\documentclass[journal=jacsat,manuscript=article,layout=twocolumn]{achemso}
\SectionNumbersOn

\usepackage{amsmath, amssymb}
\usepackage{chemformula} 
\usepackage{physics}
\usepackage[T1]{fontenc} 
\usepackage{accents} 
\usepackage{booktabs} 
\usepackage{multirow} 
\usepackage{array}
\usepackage{tikz}   
\usepackage{xcolor} 
\usepackage{siunitx} 
\newcommand{\T}{\text{T}}
\newcommand{\F}{\text{F}}
\newcommand{\E}{\mathcal{E}}
\newcommand{\I}{\mathcal{I}}
\newcommand{\B}[1]{\mathbf{#1}}
\newcommand{\argmax}{\mathop{\arg\max}}
\newcommand{\argmin}{\mathop{\arg\min}}
\newcommand{\at}[2]{{{#1}^{(#2)}}} 
\newcommand{\Bat}[2]{{\mathbf{#1}^{(#2)}}} 
\newcommand{\itell}{{\ell}}
\newcommand{\itellp}{{\ell+1}}

\newcommand{\norb}{n_{\text{orb}}} 
\newcommand{\nelec}{n_{\text{elec}}} 
\newcommand{\NFCI}{N_{\text{FCI}}}

\newcommand{\reviewerone}[1]{\textcolor{black}{#1}}
\newcommand{\reviewertwo}[1]{\textcolor{black}{#1}}

\newcommand{\revieweronecolor}{\color{black}}
\newcommand{\reviewertwocolor}{\color{black}}

\newcommand*{\mrowstyle}{}

\newcommand*{\rowstyle}[1]{
  \gdef\mrowstyle{#1}%
  \mrowstyle\ignorespaces%
}

\newcolumntype{=}{
  >{\gdef\mrowstyle{}}%
}

\newcolumntype{+}{
  >{\mrowstyle}%
}
\author{Yuejia Zhang}
\affiliation{School of Mathematical Sciences, Fudan University, Shanghai 200433, China}
\author{Weiguo Gao}
\email{wggao@fudan.edu.cn}
\affiliation{School of Mathematical Sciences, Fudan University, Shanghai 200433, China}
\alsoaffiliation{School of Data Science, Fudan University, Shanghai 200433, China}
\alsoaffiliation{Shanghai Key
Laboratory for Contemporary Applied Mathematics,
Shanghai 200433, China}
\alsoaffiliation{Key Laboratory of Computational
Physical Sciences (MOE), Shanghai 200433, China}
\author{Yingzhou Li}
\email{yingzhouli@fudan.edu.cn}
\affiliation{School of Mathematical Sciences, Fudan University, Shanghai 200433, China}
\alsoaffiliation{Shanghai Key
Laboratory for Contemporary Applied Mathematics,
Shanghai 200433, China}
\alsoaffiliation{Key Laboratory of Computational
Physical Sciences (MOE), Shanghai 200433, China}

\title{Parallel Multi-Coordinate Descent Methods for Full
Configuration Interaction}

\begin{document}

\begin{abstract}
We develop a multi-threaded parallel coordinate
descent full configuration interaction algorithm (mCDFCI), for the electronic
structure ground-state calculation in the configuration interaction
framework. The FCI problem is reformulated as an unconstrained
minimization problem, and tackled by a modified block coordinate
descent method with a deterministic compression strategy. mCDFCI
is designed to prioritize determinants based on their importance,
with block updates enabling efficient parallelization on
shared-memory, multi-core computing infrastructure.
We demonstrate the efficiency of the algorithm by computing an
accurate benchmark energy for the chromium dimer in the Ahlrichs SV
basis (48e, 42o), which explicitly includes $2.07 \times 10^9$
variational determinants. We also provide the binding curve
of the nitrogen dimer under the cc-pVQZ basis set (14e, 110o).
Benchmarks show up to $79.3\%$ parallel efficiency on 128 cores.
\end{abstract}

\section{Introduction}

Understanding the chemical properties of molecules relies on solving
the many-body time-independent electronic Schr\"{o}dinger equation.
However, traditional methods, such as density functional theory (DFT)
or coupled-cluster with single, double, and perturbative triple
excitations (CCSD(T)), often struggle to accurately describe the
electronic structure of strongly correlated systems. This limitation
is particularly evident in molecules with transition metals or
those in non-equilibrium geometries.

Full configuration interaction (FCI) provides a
numerically exact solution under a predefined basis set by describing
the wavefunction as a superposition of all possible Slater determinants.
However, FCI methods scale exponentially with the number of orbitals and
electrons, leading to the curse of dimensionality.
To leverage this challenge and apply FCI methods to large systems,
it becomes necessary to compress the wavefunction.
This can be achieved by employing different wavefunction ansatze,
such as the matrix product state (MPS) in the
density matrix renormalization group (DMRG)
method~\cite{White1999, Chan2002, Chan2011, Schollwock2011, Keller2015},
or by representing the wavefunction as a population of random particles,
as in the full configuration interaction quantum Monte Carlo (FCIQMC)
method~\cite{Booth2009, Ghanem2019, Blunt2019, Cleland2010, Yang2020}.
Another approach involves selecting important Slater determinants,
guided by the extensive sparsity of the FCI wavefunction~\cite{Eriksen2021}.
The method we describe in this paper falls into this category, which is known
as selected CI method.

A variety of selected CI methods
have been developed, starting from the earliest work in 1973 known as
Configuration Interaction using a Perturbative Selection done
Iteratively (CIPSI)~\cite{Huron1973},
to recent advancements including Adaptive
Sampling CI (ASCI)~\cite{Tubman2016},
Heat-bath CI (HCI)~\cite{Holmes2016},
Semistochastic Heat-bath CI (SHCI)~\cite{Sharma2017,Li2018},
Coordinate Descent FCI (CDFCI)~\cite{Wang2019},
Fast Randomized Iteration method for FCI (FCI-FRI)~\cite{Greene2020},
Reinforcement Learning CI (RLCI)~\cite{Goings2021},
and others.
These methods share the common iterative approach of expanding a
primary configuration space, filtering determinants based on some
importance estimate, and computing the leading eigenpair to obtain an
enhanced approximation of the wavefunction until convergence is
achieved. Typically, Epstein--Nesbet second-order perturbation theory
is further employed in the secondary space to account for the
remaining correlation that is not captured by the variational SCI
treatment. Selected CI variants significantly reduce the
computational cost of FCI by diminishing  the dimension of the
primary SCI space compared to the \(N\)-electron Hilbert space, while
they differ by having distinct selection principles and
implementations.

This paper extends the coordinate descent FCI
(CDFCI)\cite{Wang2019} method previously proposed by one
of our authors and his collaborators,
which provides selection rules from an optimization
perspective. Initially, it transforms the FCI eigenvalue problem into an
unconstrained minimization problem, with local minima corresponding
to the ground state of the system. Next, it
employs the coordinate descent method for the following
advantages: (i) The gradient of the objective function provides a
natural determinant selection rule, adding important
determinants into the variational space until it reaches the memory
limit; (ii) The special structure of the problem allows us to perform
an exact line search, accelerating the energy convergence; (iii) In
each iteration, updating only one coordinate of the optimization
vector involves only one column of the Hamiltonian matrix, avoiding
operations with the entire Hamiltonian matrix, thus reducing the
computation cost associated with unappreciative determinants.
The CDFCI method obtains the ground-state energy and
wavefunction without explicitly extracting the Hamiltonian submatrix
for direct diagonalization. This makes immense room for the storage of
the wavefunction, making it possible for larger systems and more
accurate approximations. The effective determinant selection rule and
the low storage cost are the main reasons
why CDFCI becomes a competitive FCI solver.

Although CDFCI demonstrates accelerated performance in experiments,
its parallelization capability is restricted by the inherent
sequential nature of the method. In this paper, we present a novel
algorithm to address the minimization problem, which extends update of
one coordinate to multiple coordinates per iteration. To achieve this,
the algorithm introduces an additional search dimension to enable the
exact line search. This extension not only accelerates convergence but
also opens up new possibilities for parallelization.
Benefiting from fully parallelizable coordinate updates,
our new algorithm achieves an accuracy of \(10^{-5}\) Ha for \ch{C2} and
\ch{N2} using the cc-pVDZ basis in \(10\) and \(30\) minutes respectively,
nearly twenty times faster than the single-threaded version
reported in the original CDFCI work. When compared to the
multi-threaded version of the original CDFCI that supports
parallel hashtable updates, our algorithm delivers a \(3.0 \times\) speedup.
Additionally, it computes the
ground state of all-electron \ch{Cr2} with Ahlrichs SV basis in \(5.8\) days,
matching the accuracy of the original CDFCI,
which previously required one month for the same task.

In the rest of this paper, we present the algorithm in
section~\ref{sec:method} and discuss the implementation details
in section~\ref{sec:implementation}.
In section~\ref{sec:experiments}, we demonstrate
the accuracy and the parallel efficiency of our method by applying it
to various molecules including \ch{C2}, \ch{N2} and \ch{Cr2}. The
binding curve of \ch{N2} under cc-pVQZ basis is also characterized. Finally,
we conclude and look ahead to future work in section~\ref{sec:conclusion}.

\section{Methodology and Approach}
\label{sec:method}

In this section, we begin by reviewing the reformulation of the FCI eigenvalue
problem as a nonconvex optimization problem
\cite{Wang2019, Li2019}. Following this, we describe in
detail the multi-coordinate descent methods to address the FCI problem.

\subsection{Problem Formulation}
With a complete set of one-electron spin--orbitals \(\{\chi_p\}\),
the many-body Hamiltonian operator can be expressed
using second quantization as follows:
\begin{equation}
\label{eq:second_quantized_h}
    \widehat{H} = \sum_{p,q} t_{pq} \hat{a}_{p}^\dagger \hat{a}_{q}
    + \frac{1}{2} \sum_{p,q,r,s} v_{pqrs} \hat{a}_{p}^\dagger
    \hat{a}_{q}^\dagger \hat{a}_{s} \hat{a}_{r},
\end{equation}
where \(\hat{a}_{p}^\dagger\) and \(\hat{a}_{p}\)
represent the creation and annihilation operators, respectively,
for spin--orbital $p$. The coefficients $t_{pq}$ and $v_{pqrs}$
correspond to one- and two-electron integrals.
The ground-state energy of $\widehat{H}$ is determined by
solving the time-independent Schr\"{o}dinger equation:
\begin{equation}
\label{eq:schrodinger}
\widehat{H} \ket{\Phi_0} = E_0 \ket{\Phi_0},
\end{equation}
where \(E_0\) represents the ground-state energy, i.e.,
the lowest eigenvalue of \(\widehat{H}\),
and \(\ket{\Phi_0}\) is the associated ground-state wavefunction.
We assume, without loss of generality, that $E_0$ is negative
and non-degenerate,
meaning the eigenvalues of \(\widehat{H}\) are ordered
as $E_0 < E_1 \leq E_2 \leq \dotsb$.

The full configuration interaction (FCI) method
starts from a truncated finite spin--orbital subset
\(\{\chi_p\}_{p=1}^{\norb}\), where \(\norb\) denotes the number of orbitals,
and this subset is usually obtained from a Hartree--Fock procedure.
The wavefunction is approximated as a linear combination of
Slater determinants,
\begin{equation}
\label{eq:wavefunction}
    \ket{\Phi_0} = \sum_{i=1}^{\NFCI} c_i \ket{D_i}
    = \sum_{i=1}^{\NFCI} c_i \ket{\chi_{i_1} \chi_{i_2} \dotsm \chi_{i_n}},
\end{equation}
with \(1 \le i_1 < i_2 < \dotsb < i_n \le \norb\). Here,
\(n = \nelec\) denotes the number of electrons in the system, and
\(\NFCI\) is the dimension of FCI variational space, which represents
the total number of configurations, and is of order
\(\mathcal{O}(\binom{\norb}{\nelec})\).

Following this, the Schr\"{o}dinger equation \eqref{eq:schrodinger} is
discretized into a matrix eigenvalue problem
\begin{equation}
\label{eq:schrodinger-eig}
    H\B{c} = E_0 \B{c},
\end{equation}
where \(H\) is a real symmetric \(\NFCI \times \NFCI\)
matrix, with entry \(H_{i,j} = \bra{D_i}\widehat{H}\ket{D_j}\),
and \(\B{c}\) is a vector of dimension \(\NFCI\), with entry \(c_i\).
The problem~\eqref{eq:schrodinger-eig} is known as the FCI eigenvalue problem.

The second-quantized Hamiltonian operator defined in
\eqref{eq:second_quantized_h} signifies that \(H_{i,j}\) is zero if
transforming \(\ket{D_i}\) into \(\ket{D_j}\) involves altering
more than two occupied spin--orbitals. Consequently,
\(H\) has \(\mathcal{O}(\nelec^2\norb^2)\) non-zero entries per row,
which is much smaller than \(\NFCI\) in terms of order, thus leading to an
extreme sparsity structure in the matrix \(H\).

Selected CI methods ~\cite{Huron1973,Tubman2016,Holmes2016,Sharma2017}
target a submatrix of \(H\)
and use its lowest eigenpair as an estimate to the FCI eigenvalue problem.
Now, instead of doing direct diagonalization to the submatrix of \(H\),
we consider the following optimization problem
\begin{equation}
\label{eq:optimization}
    \min_{c\in \mathbb{R}^{\NFCI}} \,
    f(\B{c}) = \min_\B{c} \,
    \|H+\B{c}\B{c}^\T\|^2_\F
\end{equation}
with the same Hamiltonian matrix \(H\) as defined above.
The gradient of the objective function is
\begin{equation}
\label{eq:gradient}
    \nabla f(\B{c}) = 4H \B{c} + 4 \left( \B{c}^\top \B{c}
    \right) \B{c}.
\end{equation}
It is shown that~\cite{Li2019} \(\pm \sqrt{-E_0} \B{v}_0 \)
are the only two local minimizers,
where \(\B{v}_0\) is the normalized eigenvector corresponding to the
smallest eigenvalue \(E_0 < 0\).
Suppose we have an estimate \(\B{\hat{c}}\)
to the solution \(\B{c^*}\) of problem~\eqref{eq:optimization},
by simply normalizing \(\B{\hat{c}}\),
we can obtain an approximate solution of the FCI eigenvalue
problem.

In this work, we follow the idea of coordinate descent,
extend one coordinate update per iteration to multiple coordinates,
and add an additional scaling factor to enable exact line search,
which is undoubtedly more beneficial to convergence. We also follow
the compression strategy in CDFCI, which improves the efficiency of
the algorithm and controls the memory footprint so that CDFCI can converge
in a subspace. Below we describe our algorithm in detail.

\subsection{Algorithm}
\label{sec:algorithm}

The multi-coordinate descent method is an iterative optimization technique
similar to the original coordinate descent method.
However, instead of modifying a single coordinate of the optimization
variable \(\B{c}\) in each iteration, we update \(k\) coordinates
simultaneously and introduce a scaling factor \(\gamma\).

Let us denote \(\Bat{c}{\itell}\) as the vector \(\B{c}\) at the
current iteration \(\itell\). We also define \(\at{\I}{\itell} =
\{\at{i_1}{\itell}, \dotsc, \at{i_k}{\itell}\} \subset
\{1, \dotsc, \NFCI\}\) as the set of \(k\) distinct coordinates chosen
for the update in this iteration.

The update rule for each component of \(\B{c}\) is given by:
\begin{equation}
\at{c_i}{\itellp} \leftarrow
\begin{cases}
    \at{\gamma}{\itell} \at{c_i}{\itell} + \at{a_i}{\itell},
    &\text{if } i \in \at{\I}{\itell}, \\
    \at{\gamma}{\itell} \at{c_i}{\itell}, &\text{otherwise.}\\
\end{cases}
\end{equation}
where \(\at{\gamma}{\itell}\) is a scaling factor,
and \(\at{a_i}{\itell}\)
is the corresponding step size.
Each entry \(\at{c_i}{\itell}\)
is initially scaled by \(\at{\gamma}{\itell}\).
For the specific coordinates \(i\) included in
the set \(\at{\I}{\itell}\), \(\at{c_i}{\itell}\) is then further
updated by adding a step size to the current value.

In matrix form, the update rule can be expressed as:
\begin{equation}
\Bat{c}{\itellp} \leftarrow \at{\gamma}{\itell} \Bat{c}{\itell}
+ \E_{\at{\I}{\itell}} \Bat{a}{\itell},
\label{eq:update-of-x}
\end{equation}
where \(\E_\I \in \mathbb{R}^{\NFCI \times k}\) is defined as
\(\E_\I = [e_{i_1}, \dotsc, e_{i_k}]\),
with \(\I = \{i_1, \dotsc, i_k\}\) and \(1 \le i_j \le \NFCI\).
In other words, \(\E_\I\) consists of columns
from the identity matrix corresponding to the selected coordinates.
Essentially, it projects the elements of
\(\Bat{a}{\itell} \in \mathbb{R}^{k}\)
into the higher-dimensional space of
\(\Bat{c}{\itell} \in \mathbb{R}^{\NFCI}\),
inserting zeros for the coordinates not included in \(\at{\I}{\itell}\).

In section~\ref{sec:coord-pick}, we discuss how we select the \(k\)
coordinates in each iteration. Next, in section~\ref{sec:coord-update},
we explain how we choose the scaling scalar \(\at{\gamma}{\itell}\)
and the stepsize vector \(\Bat{a}{\itell}\), so that they give the best
descent value in the objective function. Each local update actually
provides an energy estimate for the global problem.

Following the algorithmic design in CDFCI~\cite{Li2019},
we store an additional vector \(\B{b}\) in memory,
which approximates \(H\B{c}\) with differences only
in the coordinates that are compressed.
This will facilitate both picking coordinates and choosing the stepsize.
In section~\ref{sec:var-update}, we show how we update \(\B{b}\) and \(\B{c}\)
as well as other quantities after each iteration, where we insert a
compression strategy that keeps the memory footprint affordable.
Finally, in section~\ref{sec:pseudocode}, we provide the complete
pseudocode, along with a discussion of possible choices
for initialization and stopping criteria.

\subsubsection{Coordinate Selection}
\label{sec:coord-pick}

The selection rule of coordinates follows the largest gradients rule:
the \(k\) coordinates with largest absolute value of~\eqref{eq:gradient} are
considered. In other words, choose \(k\) coordinates, in which
\begin{equation}
    \label{eq:coord-pick0}
    i_j = \argmax_{i \ne i_1, \dotsc, i_{j-1}} \,
    \left|[\nabla f(\B{c})]_{i}\right|, \quad j = 1, \dotsc, k,
\end{equation}
where \([\nabla f(\B{c})]_{i}\) denotes the
\(i\)-th coordinate of \(\nabla f(\B{c}) = 4H\B{c} + 4
(\B{c}^\T \B{c}) \B{c}\).

Since in the algorithm design, the vector \(\B{b}\) is stored as a
compressed representation of \(H\B{c}\), at iteration \(\itell\),
the approximated gradient of the objective function can be easily
obtained by a linear combination of
\(\Bat{b}{\itell}\) and \(\Bat{c}{\itell}\).
In addition, we consider only the determinant set \(\I_H(\at{\I}{\itell})\)
which is defined as:
\begin{equation*}
    \begin{aligned}
    \I_H(\at{\I}{\itell}) := \{&1 \le i \le \NFCI: \\
    &\exists j \in \at{\I}{\itell} \text{ such that } H_{i,j} \ne 0\},
    \end{aligned}
\end{equation*}
where \(\at{\I}{\itell}\) denotes the set of coordinates selected
during the \(\itell\)-th iteration.

Therefore, the actual coordinate selection rule at iteration \(\itell\) is
\begin{equation}
    \begin{aligned}
    \label{eq:coord-pick}
    i_j^{(\itell)} = \argmax_{ \substack{
            i \in \I_H(\at{\I}{\itell-1}) \\
            i \ne i_1^{(\itell)}, \dotsc, i_{j-1}^{(\itell)}
    }} \,
    \left| 4 \at{b_{i}}{\itell} + 4 ((\Bat{c}{\itell})^\T \Bat{c}{\itell})
    \at{c_{i}}{\itell} \right|, \\
    \quad j = 1, \dotsc, k.
    \end{aligned}
\end{equation}

Our coordinate selection method is built on the concept of an active space,
which is a carefully chosen subset of the full configuration space.
This active space includes only the most important determinants,
that are expected to contribute the most to the solution.
To construct this space efficiently, we limit our consideration to
determinants that are connected to the current iterate by the
creation or annihilation of up to two electrons.
This restriction helps to avoid including determinants that are far removed
in configuration space, which would contribute minimally to the final solution.
This approach enhances computational efficiency by focusing resources
on the most relevant determinants, ensuring that the optimization
process is both effective and efficient.

\subsubsection{Optimal Scaling and Stepsize Selection}
\label{sec:coord-update}
Once we have selected \(\at{\I}{\itell}\),
the \(k\) coordinates to update,
we minimize the objective function with
respect to the scaling scalar \(\at{\gamma}{\itell}\)
and the stepsize vector \(\Bat{a}{\itell}\):
\begin{equation}
    \label{eq:obj-function}
    f(\Bat{c}{\itellp}) =
    \left\|H + \Bat{c}{\itellp}(\Bat{c}{\itellp})^\T\right\|^2_\F,
\end{equation}
where
\(\Bat{c}{\itellp} = \at{\gamma}{\itell} \Bat{c}{\itell} +
\E_{\at{\I}{\itell}} \Bat{a}{\itell}\).
In the following, we describe how we perform an exact line search in
a \((k+1)\)-dimensional subspace so that \(\at{\gamma}{\itell}\) and
\(\Bat{a}{\itell}\) are both globally optimal.

Let \(\Bat{y}{\itell} = \Bat{c}{\itell} -
\E_{\at{\I}{\itell}}(\E_{\at{\I}{\itell}})^\T \Bat{c}{\itell}\),
which sets the entries in coordinate set \(\at{\I}{\itell}\) to zeros.
Define
\begin{equation*}
    \Bat{\tilde{y}}{\itell} =
    \begin{cases}
        \Bat{y}{\itell} / \|\Bat{y}{\itell}\|,
        & \text{if } \|\Bat{y}{\itell}\|\ne 0,\\
        0, & \text{otherwise},
    \end{cases}
\end{equation*}
so that we always have \(\Bat{y}{\itell} =
\Bat{\tilde{y}}{\itell}\|\Bat{y}{\itell}\|\) regardless of
whether \(\|\Bat{y}{\itell}\| = 0\).

The update of \(\B{c}\) given in \eqref{eq:update-of-x}
can be rewritten as
\begin{equation}
    \label{eq:update-of-x-mat}
\begin{aligned}
    \Bat{c}{\itellp} &= \at{\gamma}{\itell}
    \Bat{c}{\itell} + \E_{\at{\I}{\itell}} \Bat{a}{\itell} \\
    &=
    \at{Q}{\itell} \Bat{z}{\itell},
\end{aligned}
\end{equation}
with \(\at{Q}{\itell}\) and \(\Bat{z}{\itell}\) defined as follows:
\begin{equation}
    \label{eq:define-q}
    \at{Q}{\itell} :=
    \begin{bmatrix}
        \Bat{\tilde{y}}{\itell} & \E_{\at{\I}{\itell}}
    \end{bmatrix} \in \mathbb{R}^{\NFCI \times (k+1)},
\end{equation}
\begin{equation}
    \label{eq:define-z}
    \begin{aligned}
    \Bat{z}{\itell} &:=
    \begin{bmatrix}
        \at{\gamma}{\itell} \|\Bat{y}{\itell}\| \\
        \at{\gamma}{\itell} (\E_{\at{\I}{\itell}})^\T
        \Bat{c}{\itell} + \Bat{a}{\itell}
    \end{bmatrix}.
    \end{aligned}
\end{equation}

\begin{figure*}[th]
\centering
\begin{tikzpicture}
    \def \a {-4.2}
    \def \xx {-6.5}
    \def \yy {-4.5}
    \def \zz {1.7}
    \def \inc {1.5}
    \foreach \x in {\xx,
    \yy, \yy+\inc, \yy+\inc*2, \yy+\inc*3,
    \zz, \zz+\inc, \zz+\inc*2, \zz+\inc*3}
        \draw (\x,0) rectangle (\x+0.3,\a);
    \node at (0,\a) [below = 10]
    {$\B{c}^{(\itellp)}
    \leftarrow \gamma^{(\itell)} \B{c}^{(\itell)}
    + a_1^{(\itell)} \B{e}_{i}
    + a_2^{(\itell)} \B{e}_{j}
    + a_3^{(\itell)} \B{e}_{k}
    = z_1^{(\itell)} \B{\tilde{y}}^{(\itell)}
    + z_2^{(\itell)} \B{e}_{i}
    + z_3^{(\itell)} \B{e}_{j}
    + z_4^{(\itell)} \B{e}_{k}$
    };

    \node at (-5.3, -2) {\LARGE $\leftarrow$};
    \foreach \x in {\yy, \yy+\inc, \yy+\inc*2,
                    \zz, \zz+\inc, \zz+\inc*2}
        \node at (1+\x, -2) {\Large $+$};
    \node at (1, -2) {\LARGE $=$};
    \foreach \y in {0, 1, 2, 5, 7, 8, 11, 12}
        \draw[fill=cyan] (\yy, -\y*0.3) rectangle (\yy+0.3,-\y*0.3-0.3);
    \foreach \y in {0, 1, 2, 5, 8, 12}
        \draw[fill=cyan!50] (\zz, -\y*0.3) rectangle (\zz+0.3,-\y*0.3-0.3);
    \foreach \y in {7, 11}
        \draw (\zz, -\y*0.3) rectangle (\zz+0.3,-\y*0.3-0.3);
    \foreach \y in {0, 1, 2, 4, 5, 7, 8, 11, 12}
        \draw[fill=green] (\xx, -\y*0.3) rectangle (\xx+0.3,-\y*0.3-0.3);

    \foreach \x in {\yy+\inc, \zz+\inc}
        \draw[fill=yellow] (\x, -4*0.3) rectangle (\x+0.3,-4*0.3-0.3)
                                        node[anchor=base west]{$i$};
    \foreach \x in {\yy+\inc*2, \zz+\inc*2}
        \draw[fill=yellow] (\x, -7*0.3) rectangle (\x+0.3,-7*0.3-0.3)
                                        node[anchor=base west]{$j$};
    \foreach \x in {\yy+\inc*3, \zz+\inc*3}
        \draw[fill=yellow] (\x, -11*0.3) rectangle (\x+0.3,-11*0.3-0.3)
                                         node[anchor=base west]{$k$};
    \end{tikzpicture}
    \caption{
        The update of vector \(\B{c}\) when \(k = 3\).
        After three coordinates \(i, j, k\) are picked (in yellow),
        these coordinates of \(\Bat{c}{\itell}\) are zeroed out
        and the resulting \(\Bat{y}{\itell}\) is normalized to
        \(\Bat{\tilde{y}}{\itell}\) (in light blue).
        The coefficients \(a^{(\itell)}_i\) and \(z^{(\itell)}_i\)
        represent the \(i\)-th element of vectors \(\Bat{a}{\itell}\)
        and \(\Bat{z}{\itell}\) respectively.
    }
    \label{fig:update-c}
\end{figure*}
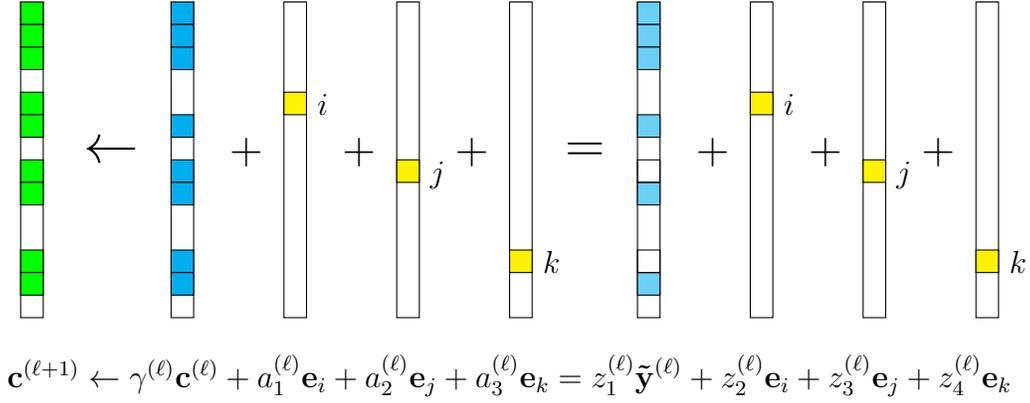

Figure~\ref{fig:update-c} demonstrates how the vector \(\B{c}\) is
updated in each iteration.

If \(\|\Bat{y}{\itell}\|\ne 0\),
\(\at{Q}{\itell}\) is an orthogonal matrix and an orthogonal basis for
the \((k+1)\)-dimensional search space.
If \(\|\Bat{y}{\itell}\|= 0\), \((\at{Q}{\itell})^\T\at{Q}{\itell} =
\begin{bmatrix} 0 & \\ & I_k \end{bmatrix}\), which means
the last \(k\) columns of \(\at{Q}{\itell}\) form an orthogonal basis for
the \(k\)-dimensional search space.

Now consider split the optimization problem~\eqref{eq:obj-function}
into two parts as follows:
\begin{equation}
    \label{eq:obj-function-transformed}
    \begin{aligned}
    &f(\Bat{c}{\itellp}) =
    \left\|H +\Bat{c}{\itellp}{(\Bat{c}{\itellp})}^\T\right\|^2_\F \\
    &= \left\|\left(\at{Q}{\itell}(\at{Q}{\itell})^\T\right)
    \left(H +\Bat{c}{\itellp}{(\Bat{c}{\itellp})}^\T\right)\right\|^2_\F \\
    &+ \left\|\left(I - \at{Q}{\itell}(\at{Q}{\itell})^\T\right)
    \left(H +\Bat{c}{\itellp}{(\Bat{c}{\itellp})}^\T\right)\right\|^2_\F \\
    &= \left\|(\at{Q}{\itell})^\T \left(H + \Bat{c}{\itellp}
       {(\Bat{c}{\itellp})}^\T\right)\at{Q}{\itell}\right\|^2_\F \\
    &+ \left\|\left(I - \at{Q}{\itell}(\at{Q}{\itell})^\T\right)H
       \right\|^2_\F \\
    &=\left\|(\at{Q}{\itell})^\T H \at{Q}{\itell}
    + \Bat{z}{\itell} (\Bat{z}{\itell})^\T \right\|^2_\F + \text{constant}.
    \end{aligned}
\end{equation}
The third equality follows from the fact that
\(\left(I - \at{Q}{\itell}(\at{Q}{\itell})^\T\right)
\Bat{c}{\itellp} = 0\),
and the last equality is derived from equation~\eqref{eq:update-of-x-mat}.
In addition, the term
\(\left\|\left(I - \at{Q}{\itell}(\at{Q}{\itell})^\T\right)H\right\|^2_\F\)
is considered as a constant because it does not depend on
\(\Bat{z}{\itell}\), and therefore neither on \(\at{\gamma}{\itell}\)
nor \(\Bat{a}{\itell}\).

We choose \(\Bat{z}{\itell}\) to be the optimal solution to the local
minimization problem
\begin{equation}
    \label{eq:optimization-z}
    \begin{aligned}
    \Bat{z}{\itell} &= \argmin_{\B{z}\in \mathbb{R}^{k+1}} \,
    f(\B{z}) \\
    & = \argmin_{\B{z}\in \mathbb{R}^{k+1}} \,
    \left\|(\at{Q}{\itell})^\T H \at{Q}{\itell}
    + \B{z} \B{z}^\T \right\|^2_\F.
    \end{aligned}
\end{equation}
The minimum is attained when~\cite{Li2019}
\begin{equation}
    \label{eq:z}
    \Bat{z}{\itell} = \pm \sqrt{-\at{\lambda}{\itell}}
    \B{v}_{\at{\lambda}{\itell}},
\end{equation}
where
\(\at{\lambda}{\itell} := \lambda_{\min}((\at{Q}{\itell})^\T H\at{Q}{\itell})\)
is the minimal eigenvalue of matrix \((\at{Q}{\itell})^\T H \at{Q}{\itell}\)
and \(\B{v}_{\at{\lambda}{\itell}}\) is
the corresponding eigenvector. The eigenvalue
\(\at{\lambda}{\itell}\) is guaranteed to be negative
as long as \((\Bat{c}{1})^\T H \Bat{c}{1} < 0\)
(see Appendix~\ref{app:proof} for proof and section~\ref{sec:pseudocode} for
the discussion of initialization).

The problem \eqref{eq:optimization-z} consists of two parts:
assembling the \((\at{Q}{\itell})^\T H \at{Q}{\itell}\) matrix
and calculating its leading eigenpair.
To start, we see that,
\begin{equation}
    \label{eq:submatrix}
    \begin{aligned}
        &(\at{Q}{\itell})^\T H \at{Q}{\itell} \\
& =
\begin{bmatrix} (\Bat{\tilde{y}}{\itell})^\T \\
    \E_{\at{\I}{\itell}}^\T \end{bmatrix} H
\begin{bmatrix}\Bat{\tilde{y}}{\itell} &
    \E_{\at{\I}{\itell}}\end{bmatrix} \\
&=
\begin{bmatrix} (\Bat{\tilde{y}}{\itell})^\T H \Bat{\tilde{y}}{\itell} &
(\Bat{\tilde{y}}{\itell})^\T H \E_{\at{\I}{\itell}} \\
 \E_{\at{\I}{\itell}}^\T H \Bat{\tilde{y}}{\itell} &
 \E_{\at{\I}{\itell}}^\T H \E_{\at{\I}{\itell}}
\end{bmatrix}.
\end{aligned} \end{equation}
If \(\|\Bat{y}{\itell}\|\ne 0\),
each block is computed as described below.
\begin{enumerate}
    \item \((\Bat{\tilde{y}}{\itell})^\T H \Bat{\tilde{y}}{\itell}
          \in \mathbb{R}\): it is evaluated by
          \(\frac{1}{\|\Bat{y}{\itell}\|^2}
          \Bigl((\Bat{c}{\itell})^\T\Bat{b}{\itell}
                - 2\sum_{i \in \at{\I}{\itell}}
                \at{c_i}{\itell} \at{b_i}{\itell}
                + \sum_{i,j \in \at{\I}{\itell}}
                \at{c_i}{\itell} H_{i,j} \at{c_j}{\itell}\Bigr).\)
    \item \(((\Bat{\tilde{y}}{\itell})^\T H \E_{\at{\I}{\itell}})^\T =
          \E_{\at{\I}{\itell}}^\T H \Bat{\tilde{y}}{\itell} \in \mathbb{R}^k\):
          its \(i\)-th entry is evaluated by
          \(\frac{1}{\|\Bat{y}{\itell}\|}
          \Bigl(\at{b_i}{\itell}
          - \sum_{j \in \at{\I}{\itell}} H_{i,j} \at{c_j}{\itell}\Bigr).\)
    \item \( \E_{\at{\I}{\itell}}^\T H \E_{\at{\I}{\itell}}
             \in \mathbb{R}^{k\times k}\):
          It is the submatrix of \(H\) with index set \(\at{\I}{\itell}\).
          For \(i, j \in \at{\I}{\itell}\), the entry \(H_{i,j}\)
          is evaluated on the fly by
          \(H_{i,j} = \bra{D_i}\widehat{H}\ket{D_j}\).
\end{enumerate}

In the other case when \(\|\Bat{y}{\itell}\| = 0\), only the matrix block
\( \E_{\at{\I}{\itell}}^\T H \E_{\at{\I}{\itell}}\) is non-zero.

When \(k\) is not too large, the smallest eigenpair of
\((\at{Q}{\itell})^\T H \at{Q}{\itell}\) can be easily
retrieved by default LAPACK eigensolvers~\cite{MRRR} for selected eigenvalues.
Once this is complete,
\(\Bat{z}{\itell}\) is computed according to equation~\eqref{eq:z}.
By equation~\eqref{eq:define-z}, we have
\begin{equation}
    \label{eq:coord-update-gamma}
    \at{\gamma}{\itell} =
    \begin{cases}
        \frac{\at{z_1}{\itell}}{\|\Bat{y}{\itell}\|}, &
        \text{if } \|\Bat{y}{\itell}\| \ne 0, \\
        1, &  \text{otherwise,}
    \end{cases}
\end{equation}
and
\begin{equation}
    \label{eq:coord-update-a}
    \Bat{a}{\itell} = \Bat{z_{2:}}{\itell} -
    \at{\gamma}{\itell} (\E_{\at{\I}{\itell}})^\T
    \Bat{c}{\itell},
\end{equation}
where \(\at{z_1}{\itell}\) denotes the first entry of \(\B{z}\),
and \(\Bat{z_{2:}}{\itell} \in \mathbb{R}^k\) the rest of the vector.
Note that for the case of \(\|\Bat{y}{\itell}\| = 0\),
\(\at{z_1}{\itell}\) would always be zero, and
\(\at{\gamma}{\itell}\) can be any value.
For simplicity, we just set \(\at{\gamma}{\itell} = 1\).

\subsubsection{Variable Update and Coefficient Compression}
\label{sec:var-update}

After obtaining the scaling factor \(\at{\gamma}{\itell}\)
and the stepsize \(\Bat{a}{\itell}\) from the procedure described in
section~\ref{sec:coord-update}, vectors \(\B{c}\) and \(\B{b}\)
are updated by
\begin{equation}
    \label{eq:update-c}
    \Bat{c}{\itellp} \leftarrow \at{\gamma}{\itell} \Bat{c}{\itell}
    + \E_{\at{\I}{\itell}} \Bat{a}{\itell},
\end{equation}
and
\begin{equation}
    \label{eq:update-b}
    \Bat{b}{\itellp} \leftarrow \at{\gamma}{\itell}
    \Bat{b}{\itell} + H \E_{\at{\I}{\itell}} \Bat{a}{\itell}.
\end{equation}
It should be noted that in the actual implementation,
\(\Bat{c}{\itellp}\) is not updated exactly as shown above.
Instead, we store the scaling factor \(\gamma\) and modify only
the relevant entries of \(\Bat{c}{\itell}\), while most entries
remain unchanged. For a detailed description of the implementation,
please refer to section~\ref{sec:implementation}.

Through the above update rules~\eqref{eq:update-c} and~\eqref{eq:update-b},
the number of non-zero elements in \(\B{c}\) and \(\B{b}\) will continue
to grow. Specifically, the growth rate for \(\B{b}\) is much faster than
\(\B{c}\) in the beginning,
because the number of entries involved in updating \(\B{b}\)
in~\eqref{eq:update-b} is much greater than updating c in~\eqref{eq:update-c}.
This process can be seen as continuously adding determinants to
the variational space, the selected subspace of the full CI space
where we approximate the ground state.
If we continue to iterate without compression,
the variational space will eventually expand to the full CI space,
which is obviously not feasible for large systems. To this end,
we propose a compression strategy that targets \(\B{b}\).
The update of \(\Bat{b}{\itellp}\) for a certain coordinate \(i\)
will be discarded if both of the following conditions are satisfied:
(i) if \(\at{b_i}{\itell} = 0\), indicating that the determinant \(\ket{D_i}\)
has not been selected before; (ii) \(\at{\Delta b_i}{\itellp} \le \tau\),
with \(\Bat{\Delta b}{\itellp} := H
\E_{\at{\I}{\itellp}} \Bat{a}{\itellp}\), meaning that the update is
relatively small. For any determinant \(\ket{D_i}\) deemed unimportant,
as long as both \(b_i\) and \(c_i\) remain zero,
it will not be selected during the coordinate selection step. Thus,
the size of variational space is controlled, and the wavefunction vector
will converge into a subspace of the full CI space, consisting
only of limited number of important determinants.

Notably, despite compressing the vector \(\B{b}\),
\(b_i\) remains accurate for every determinant \(\ket{D_i}\) with
nonzero coefficients in \(\B{c}\), meaning it was selected at least once
during the coordinate selection step. This is achieved by recalculating
\(\at{b_i}{\itellp}\) for each selected determinant \(\ket{D_i}\)
selected at the \(\itell\)-th iteration, using the following formula:
\begin{equation}
    \at{b_i}{\itellp} = H_{i,:} \Bat{c}{\itellp}
    = \sum_{j \in \I_H(i)} H_{i,j} \at{c_j}{\itellp}.
    \label{eq:recalculate-b}
\end{equation}

Once a determinant \(\ket{D_i}\) is selected and recalculated,
updates to \(b_i\) will not be discarded again according to the
compression rule. Therefore, \(\B{b}\) is only compressed
for determinants that have not yet been added to the variational space.
This ensures that the Rayleigh quotient
\((\B{c}^\T H\B{c})/(\B{c}^\T \B{c})\),
is accurately maintained in each iteration by
\[\frac{(\Bat{c}{\itell})^\T H \Bat{c}{\itell}}
{(\Bat{c}{\itell})^\T \Bat{c}{\itell}} =
\frac{(\Bat{c}{\itell})^\T \Bat{b}{\itell}}
{(\Bat{c}{\itell})^\T \Bat{c}{\itell}}
,\]
since the compressed terms \(b_i\) always have
associated \(c_i = 0\) and therefore do not contribute to the sum.

Both the numerator and denominator are stored in memory, and
updated in each iteration
via the following straight-forward update rules:
\begin{equation}
    \label{eq:update-cc}
    \begin{aligned}
    (\Bat{c}{\itellp})^\T \Bat{c}{\itellp}
    & \leftarrow (\at{\gamma}{\itell})^2
        (\Bat{c}{\itell})^\T \Bat{c}{\itell} \\
    & + 2 \at{\gamma}{\itell} (\Bat{a}{\itell})^\T
        (\E_{\at{\I}{\itell}} \Bat{c}{\itell}) \\
    & + (\Bat{a}{\itell})^\T \Bat{a}{\itell},
    \end{aligned}
\end{equation}
and
\begin{equation}
    \label{eq:update-cb}
    \begin{aligned}
    (\Bat{c}{\itellp})^\T \Bat{b}{\itellp}
    & \leftarrow (\at{\gamma}{\itell})^2
        (\Bat{c}{\itell})^\T \Bat{b}{\itell} \\
    & + 2 \at{\gamma}{\itell} (\Bat{a}{\itell})^\T
        (\E_{\at{\I}{\itell}} \Bat{b}{\itell}) \\
    & + (\Bat{a}{\itell})^\T H \Bat{a}{\itell}.
    \end{aligned}
\end{equation}

\subsubsection{Proposed Multi-Coordinate Descent FCI Method}
\label{sec:pseudocode}

The following outlines the complete procedure
for the Multi-Coordinate Descent FCI method.

\begin{enumerate}
    \item Initialize \(\at{\I}{0}\), the set of determinants
    pre-selected before computation begins.
    In our case, we choose \(\at{\I}{0} = \{\ket{D_\text{HF}}\}\),
    which introduces only one non-zero entry in \(\Bat{c}{1}\),
    simplifying the computation of \(\Bat{b}{1}\).
    Set \(\itell = 1\), initialize \(\Bat{c}{1}\), and
    compute \(\Bat{b}{1} = H \Bat{c}{1}\).
    Other initial vectors may be chosen,
    provided that \((\Bat{c}{1})^\T H \Bat{c}{1} < 0\) holds.
    \item Select the index set \(\at{\I}{\itell} = \{\at{i_1}{\itell}, \dotsc,
    \at{i_k}{\itell}\}\) according to~\eqref{eq:coord-pick}.
    \item Evaluate the matrix \((\at{Q}{\itell})^\T H \at{Q}{\itell}\)
    following~\eqref{eq:submatrix}. Compute its lowest eigenvalue
    and corresponding eigenvector \(\Bat{z}{\itell}\), and then obtain
    the scaling factor \(\at{\gamma}{\itell}\)
    using~\eqref{eq:coord-update-gamma}
    and the stepsize vector \(\Bat{a}{\itell}\) via~\eqref{eq:coord-update-a}.
    \item Update \(\Bat{c}{\itellp}\), \(\Bat{b}{\itellp}\),
    \((\Bat{c}{\itellp})^\T \Bat{c}{\itellp}\) and
    \((\Bat{c}{\itellp})^\T \Bat{b}{\itellp}\)
    according to~\eqref{eq:update-c}, \eqref{eq:update-b},
    \eqref{eq:update-cc} and~\eqref{eq:update-cb}.
    Use the compression technique based on a predefined threshold \(\tau\).
    Recalculate \(\at{b_i}{\itellp}\) for \(i \in \at{\I}{\itell}\)
    as in~\eqref{eq:recalculate-b}.
    \item Repeat steps 2--4 with \(\itell \leftarrow \itellp \)
    until either a stopping criterion
    or the maximum iteration number is reached.
    In our implementation, we adopt the stopping criterion
    proposed in OptOrbFCI~\cite{Li2020}.
    The moving average \(S\) of the stepsize norm is tracked using
    a decay factor of \(\alpha = 0.99\),
    \[S^\itellp \leftarrow \alpha S^\itell + (1-\alpha) \|\Bat{a}{\itellp}\|.\]
    The algorithm halts when \(S^\itellp\) falls below a specified
    tolerance level.
    Other stopping criteria, such as monitoring the change
    of the Rayleigh quotient or the ratio of the number of
    nonzero coefficients in \(\B{b}\) and \(\B{c}\)~\cite{Wang2019},
    may also be applied.
\end{enumerate}

\section{Computational Implementation and Complexity Analysis}
\label{sec:implementation}

We will now provide implementation details of the algorithm,
focusing on computationally expensive components and numerical
stability issues, while conducting complexity analysis in the same time.
In the following analysis, we denote \(N_H = \max_{i}\I_H(i)\)
as the maximum number of nonzero entries in columns of the Hamiltonian matrix,
which scales as \(\mathcal{O}(\nelec^2 \norb^2)\).
We claim that the per-iteration complexity of our algorithm is
\(\mathcal{O}(kN_H + k^3)\), and \(\mathcal{O}(N_H + k^3)\) for each thread
with shared memory parallelism.

\begin{table*}[th]
    \centering
\begin{tabular}{ccc}
    \toprule
    \textbf{Step} & \textbf{Complexity per thread} & \textbf{Complexity in total}\\
    \midrule
    Coordinate pick & \(\mathcal{O}(N_H + k^2)\) &
    \(\mathcal{O}(kN_H + k^2)\) \\
    Subproblem solve & \(\mathcal{O}(k^3)\) & \(\mathcal{O}(k^3)\) \\
    Variable update & \(\mathcal{O}(N_H)\) & \(\mathcal{O}(kN_H)\) \\
    \bottomrule
\end{tabular}
    \caption{Computational cost for each step per thread and in total.
    \(N_H\) is defined as the maximum number of nonzero entries
    in any column of the Hamiltonian matrix, i.e.,
    \(N_H = \max_{i}\I_H(i)\).}
    \label{tab:complexity}
\end{table*}

The sparse vectors \(\B{c}\) and \(\B{b}\) are stored in a hashtable
that allows parallel read/write operations~\cite{libcuckoo}
as in previous work~\cite{Wang2019}.
For each entry of the hash table, its key
is the Slater determinant \(\ket{D_i}\)
encoded by an \(\norb\)-bit binary string,
and its value is a pair of double floating-point
numbers \(c_i\) and \(b_i\).
Three additional scalars are also stored in memory and updated in each
iteration: the inner products \(\B{c}^\T \B{c}\), \(\B{c}^\T \B{b}\)
and the scaling factor \(\gamma\).
All the three quantities are stored in quadruple-precision floating
point format to mitigate accumulated numerical errors,
ensuring accuracy unless the number of iterations exceeds
\(10^{16}\)~\cite{Wang2019}.

Note that the Hamiltonian matrix is never stored in memory: indeed,
we evaluate each Hamiltonian entry on-the-fly. Different from other
selected CI methods, this significantly reduces the storage cost of
our algorithm, and allows us to utilize all available storage capacity
to store the ground-state coefficient vector \(\B{c}\)
and its associated \(\B{b} = H\B{c}\) related to the residual.

Our implementation utilizes shared memory parallelism based on OpenMP.
Specifically, we enable \(k\) threads where \(k\) is the number of
coordinates we choose to update in each iteration.

In the coordinate pick stage (described in section~\ref{sec:coord-pick}),
each thread handling determinant \(\ket{D_i}\) examines determinants
in \(\I_H(i)\) to select \(k\) determinants with the
largest value of \([\nabla f(\B{c})]_{i}\). This step has a complexity of
\(\mathcal{O}(N_H)\) for each thread, and \(\mathcal{O}(kN_H)\) in total.
After this, the main thread selects the \(k\) largest determinants
from \(k^2\) gathered from all the threads.

In the subproblem solve stage (described in section~\ref{sec:coord-update}),
we begin by constructing the \((k+1)\)-dimensional
matrix \((\at{Q}{\itell})^\T H \at{Q}{\itell}\) in~\eqref{eq:submatrix}.
It is evident that computing
\((\Bat{\tilde{y}}{\itell})^\T H \Bat{\tilde{y}}{\itell}\)
and \(\E_{\at{\I}{\itell}}^\T H \Bat{\tilde{y}}{\itell}\)
involves only \(\mathcal{O}(k^2)\) cost,
whereas for \( \E_{\at{\I}{\itell}}^\T H \E_{\at{\I}{\itell}}\),
we need to evaluate \(k^2\) Hamiltonian entries, each with complexity
\(\mathcal{O}(\norb)\).
For the subsequent eigenvalue problem, we want to retrieve the
smallest eigenvalue and its corresponding eigenvector of a
\((k+1)\)-dimensional symmetric matrix. A robust eigensolver
like MRRR~\cite{MRRR}
yields such eigenpair at a cost of \(\mathcal{O}(k^3+Tk^2)\) where
\(T\) is the number of iteration steps, which is usually comparable to
\(k\).

Unfortunately, MRRR only guarantees that the extreme eigenvalue is converged
to double precision, without guaranteeing the accuracy of each entry in the
eigenvector. This could be problematic because the magnitudes of coordinates
in the eigenvector can vary widely, and the small values of the stepsize
vector might only have an accuracy of \(10^{-8}\). To address this issue,
we employ one step of inverse iteration conducted in quadruple precision
to improve the accuracy of eigenvector, which again has a complexity of
\(\mathcal{O}(k^3)\).

Finally, during the variable update stage (described in
section~\ref{sec:var-update}), each thread is assigned to calculate
a local update of \(\B{b}\) in the form \(\B{\Delta b}^i = H_{:,i} a_i\).
Compression occurs at the thread level, skipping updates of coordinate
\(j\) if \(b_j = 0\) and \(\B{\Delta b}^i_j \le \tau\).
In this process, each thread constructs the \(i\)-th column of \(H\),
\(H_{:,i}\), which is subsequently used to recalculate \(\at{b_i}{\itellp}\)
as described in~\eqref{eq:recalculate-b}.
The complexity of the variable update step remains
\(\mathcal{O}(N_H)\) per thread, resulting in a total complexity of
\(\mathcal{O}(kN_H)\). The prefactor is dominated by the computation cost
of a single Hamiltonian entry, which scales with \(\mathcal{O}(\norb)\).

\reviewerone{In the proposed algorithm, two synchronization points occur in each iteration: one after coordinate selection and another after the variable update. Specifically, Update \eqref{eq:recalculate-b} is performed sequentially following the parallel execution of \eqref{eq:update-c} and \eqref{eq:update-b}, ensuring all updates are completed and maintaining data consistency across threads.}

Table~\ref{tab:complexity} gives a summary of computational cost for each
step per thread and in total.

\section{Numerical Experiments}
\label{sec:experiments}
In this section, we conduct a series of numerical experiments to demonstrate
the efficiency of the proposed algorithm -- multi-coordinate descent
FCI method (mCDFCI). First, we compute the ground-state energy of \ch{N2}
using three different basis sets: cc-pVDZ, cc-pVTZ, and cc-pVQZ.
As the number of orbitals increases, the calculated energies become
more accurate, but the computational cost also rises.
We test the computations using different numbers of cores and demonstrate the
strong parallel scaling for big systems.
Subsequently, we compare the convergence curve and computation
time with the original CDFCI method, as well as other methods,
namely SHCI, DMRG and \reviewertwo{iS-FCIQMC (initiator semi-stochastic FCIQMC)}.
After that, we turn our eyes on a bigger system,
the chromium dimer \ch{Cr2} under the Ahlrichs SV basis at \(r = 1.5\)\r{A},
with \(48\) electrons and \(84\) spin--orbitals.
This is a well-known multi-reference system because of the spin coupling of the
\(12\) valence electrons. Finally, we benchmark the binding curve of nitrogen
dimer under the cc-pVQZ basis, which consists of \(220\) spin--orbitals.

In all experiments, the orbitals and integrals are calculated via restricted
Hartree--Fock (RHF) in PSI4 package~\cite{psi4} version 1.8. All energies are
reported in the unit of Hartree (Ha).

The program is compiled using the GNU g++ compiler version 11.4.0 with the --O3
optimization flag and its native OpenMP support. It is linked against
LAPACK version 3.11.0 and OpenBLAS version 0.3.26 for numerical linear
algebra computations.

\subsection{Scalability and Performance Testing}
\label{sec:scalability}
\begin{table*}[th]
    \centering
    \begin{tabular}{lcccccc}
        \toprule
        \textbf{Basis Set} & \textbf{Electrons} & \textbf{Spin Orbitals}  & \textbf{Threshold $\tau$} & \textbf{$N_H$} \\
        \midrule
        cc-pVDZ & 14 & 56  & $5 \times 10^{-7}$ & $< 4 \times 10^{3}$ \\
        cc-pVTZ & 14 & 120 & $5 \times 10^{-6}$ & $< 3 \times 10^{4}$ \\
        cc-pVQZ & 14 & 220 & $5 \times 10^{-5}$ & $< 8 \times 10^{4}$ \\
        \bottomrule
    \end{tabular}
    \caption{Description of the basis sets, configurations, thresholds,
    and average nonzero per column used for \ch{N2} in
    section~\ref{sec:scalability}.
    \(N_H\) is defined as the maximum number of nonzero entries
    in any column of the Hamiltonian matrix, i.e.,
    \(N_H = \max_{i}\I_H(i)\).}
    \label{tab:basis_sets}
\end{table*}

\begin{figure*}[th]
    \centering
    \includegraphics[width=\textwidth]{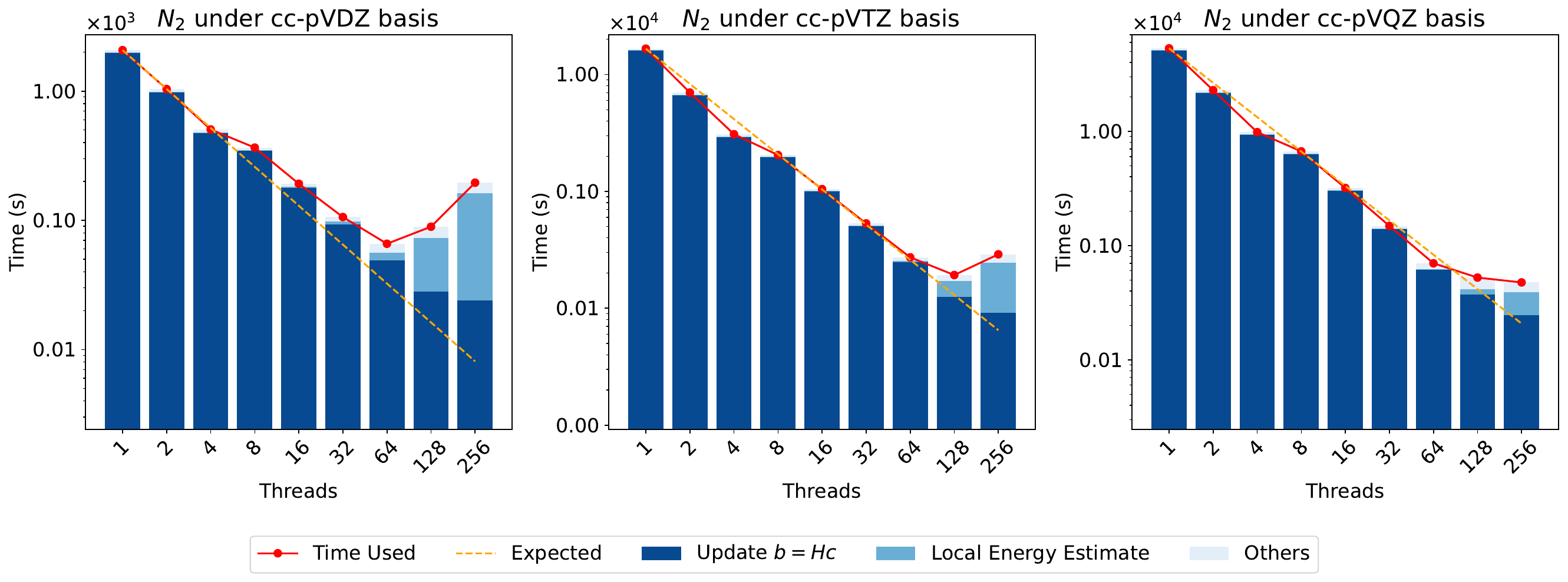}
    \caption{
        Comparison of runtime for the \ch{N2}
        molecule with different basis sets (cc-pVDZ, cc-pVTZ, and cc-pVQZ)
        using various numbers of threads. Each subplot represents a different
        basis set. The bar charts show the time in seconds spent on specific
        computational steps, including updating \( \B{b} = H\B{c} \), local
        energy estimation, and other operations.
        The red dots on the line plot show the total time usage,
        while the yellow dashed line represents the ideal scaling behavior
        (expected performance improvement with increased threads).
    }
    \label{fig:scalability}
\end{figure*}

The \ch{N2} molecule is considered a correlated system due to the
significant role of electron correlation effects in accurately
describing its electronic structure. We compute its ground-state energy
using three different basis sets: cc-pVDZ (14e, 28o), cc-pVTZ (14e, 60o),
and cc-pVQZ (14e, 110o). A detailed description of the system and the
configurations used when running the CDFCI program is provided in
Table~\ref{tab:basis_sets}. The experiments in this section were
conducted on a system equipped with two AMD EPYC 9754 128-core
processors and 1.5 TB of memory.

In the multi-threaded implementation of CDFCI (mCDFCI),
we define \emph{effective iterations} as the product of
the number of iterations and the number of determinants updated,
ensuring that mCDFCI performs an equivalent workload regardless
of the number of threads.
Notably, in our experiments,
while different values of \(k\)  produce distinct energy descent
trajectories, the energy values after the same number of effective
iterations remain consistent,
irrespective of the number of threads used.
Specifically, for the cc-pVDZ basis,
the maximum difference between energy values was \(4.84 \times 10^{-7}\) Ha,
while for cc-pVTZ and cc-pVQZ, the differences were
\(2.66 \times 10^{-6}\) Ha and \(9.09 \times 10^{-6}\) Ha, respectively.
All of these differences are smaller than \(10^{-5}\) Ha,
confirming the robustness and stability of the method.

Figure~\ref{fig:scalability} shows the runtime
of mCDFCI over \(1024\)K effective iterations,
evaluated across a range of thread counts from \(1\) to \(256\).
\reviewerone{The detailed data supporting these results are provided in
Appendix~\ref{app:scalability}, where the exact runtime of
each component is presented, along with an in-depth analysis
of the parallel capabilities.}
In every case, the dominant computational cost arises
from the update of the vector \( \B{b} = H\B{c} \).
\reviewerone{However, the time for the local energy estimation,
which is not parallelized, increases as the thread count grows.
This behavior can be attributed to the increasing impact of the \(k^3\) term in the eigenvalue computation as \(k\) becomes larger.}
This increase is particularly noticeable for smaller systems,
such as cc-pVDZ, where the computational time grows significantly
beyond \(64\) threads.

For larger systems, strong parallel scalability is observed
up to \(128\) threads, with instances of superlinear speedup.
This performance gain is attributed to the hashtable we use,
which is specifically designed and optimized for multi-threaded workloads.
When focusing solely on the time taken to update the hash table
and disregarding the local energy estimation,
it is clear that the update of the \(\B{b}\) vector is highly parallelizable.

\subsection{Numerical Results with \ch{C2} and \ch{N2}}

In this section, we compare the performance of our method
with other FCI methods: SHCI, DMRG and \reviewertwo{iS-FCIQMC},
all with parallel support and using the same amount of computing resources.
We test all the algorithms on two systems:
\ch{C2} and \ch{N2} under the cc-pVDZ basis sets.
The original CDFCI program~\cite{cdfci} which supports OpenMP acceleration
by parallelizing the update \(\B{b} = H\B{c}\), is also tested here
and renamed as sCDFCI (single-coordinate descent FCI)
for comparison with our new method mCDFCI (multi-coordinate descent FCI).
SHCI, DMRG and \reviewertwo{iS-FCIQMC} calculations
were performed using the Arrow code~\cite{shci}, the Block2 code~\cite{dmrg}
and the NECI code~\cite{fciqmc}, respectively.
All programs were compiled using the GNU g++ compiler (version 11.4.0)
with the --O3 optimization flag. Native support for OpenMP was utilized,
and MPI support was provided by MPICH (version 4.2.0).
All programs were restricted to use \(64\) cores.
CDFCI, SHCI, and DMRG were executed with 64 threads each, while
\reviewertwo{iS-FCIQMC}, which supports only MPI parallelization,
was executed with 64 MPI processes.
Apart from the parallelization configurations enabled in our study,
we used the same experimental settings as described in~\citet{Wang2019}.

Specifically, SHCI was run with different thresholds \(\varepsilon_1\),
DMRG with varying maximum bond dimensions \(M\),
and \reviewertwo{iS-FCIQMC} with different numbers of walkers \(m\).
\reviewertwo{For iS-FCIQMC, the preconditioned heat-bath (PCHB)
excitation generator and the semi-stochastic method
(with \(\num{1000}\) determinants in the deterministic space) were employed.}
The initiator threshold was set to \(3.0\) and the number of iterations
was set to $\num{30000}$ to ensure convergence by observation of clear
plateaus in the diagnostic plots.
The Hartree--Fock state was used as the initial wavefunction for
all algorithms except DMRG, which has its own warm-up algorithm.
In all cases, the energy is reported without any perturbation
or extrapolation postcalculations. Variational energy is reported
for CDFCI, SHCI and DMRG, \reviewertwo{while average projected energy
is reported for iS-FCIQMC, along with stochastic error estimates
obtained through additional blocking analysis.}
\reviewerone{The absolute error is defined as
the absolute difference between the
reported energy and the reference ground-state energy.}

\begin{table*}[ht]
    \centering
    \begin{tabular}{=c =l|+r +r +r|+r +r}
        \toprule
        \multirow{2}{*}{\textbf{Algorithm}} & \multirow{2}{*}{\textbf{Parameter}}
        & \multirow{2}{*}{\textbf{Absolute Error}} & \multirow{2}{*}{\textbf{Mem(GB)}}
        & \multirow{2}{*}{\textbf{Time(s)}} & \multicolumn{2}{c}{\textbf{mCDFCI}} \\
        & & & & & \textbf{Time(s)} & \textbf{Ratio} \\
        \midrule
\multirow{5}{*}{mCDFCI}
& \multirow{5}{*}{$\tau = 3.0 \times 10^{-8}$}
  & $1.0 \times 10^{-2}$ & $1.5$ & $3.8$   & - & \\
& & $1.0 \times 10^{-3}$ & $6.0$ & $17.2$   & - & \\
& & $1.0 \times 10^{-4}$ & $24.0$ & $123.4$   & - & \\
& & $1.0 \times 10^{-5}$ & $48.0$ & $603.7$   & - & \\
& & $1.0 \times 10^{-6}$ & $48.0$ & $2332.3$   & - & \\
\midrule
\multirow{5}{*}{sCDFCI}
& \multirow{5}{*}{$\varepsilon = 3.0 \times 10^{-8}$}
  & $1.0 \times 10^{-2}$ & $3.0$ & $14.7$ &  $3.8$ & $3.87\times$\\
& & $1.0 \times 10^{-3}$ & $6.0$ & $55.0$ &  $17.2$ & $3.20\times$\\
& & $1.0 \times 10^{-4}$ & $24.0$ & $374.3$ &  $123.4$ & $3.03\times$\\
& & $1.0 \times 10^{-5}$ & $48.0$ & $1779.7$ &  $603.7$ & $2.95\times$\\
& & $1.0 \times 10^{-6}$ & $48.0$ & $6906.0$ &  $2332.3$ & $2.96\times$\\
\midrule
\multirow{5}{*}{SHCI}
& $\varepsilon_1 = 1.0\times 10^{-4}$
& $1.5\times 10^{-3}$ & $16.1$ & $3.3$ &  $14.0$ & $0.23\times$ \\
& $\varepsilon_1 = 5.0\times 10^{-5}$
& $7.6\times 10^{-4}$ & $17.5$ & $7.3$ &  $23.7$ & $0.31\times$ \\
& $\varepsilon_1 = 1.0\times 10^{-5}$
& $1.1\times 10^{-4}$ & $32.6$ & $53.9$ &  $113.2$ & $0.48\times$ \\
& $\varepsilon_1 = 5.0\times 10^{-6}$
& $4.6\times 10^{-5}$ & $54.2$ & $126.0$ &  $219.4$ & $0.57\times$ \\
& $\varepsilon_1 = 1.0\times 10^{-6}$
& $4.7\times 10^{-6}$ & $238.1$ & $845.0$ &  $954.1$ & $0.89\times$ \\
\midrule
\multirow{4}{*}{DMRG}
& $\max M = 500$
& $6.9\times 10^{-4}$ & $0.2$ & $106.9$ &  $23.7$ & $4.52\times$ \\
& $\max M = 1000$
& $1.5\times 10^{-4}$ & $0.8$ & $345.2$ &  $92.7$ & $3.72\times$ \\
& $\max M = 2000$
& $1.9\times 10^{-5}$ & $3.1$ & $1130.1$ &  $394.9$ & $2.86\times$ \\
& $\max M = 4000$
& $2.1\times 10^{-6}$ & $11.5$ & $3848.3$ &  $1549.2$ & $2.48\times$ \\
\midrule
\multirow{5}{*}{\reviewertwo{iS-FCIQMC}}
& $m = 10000$
\rowstyle{\reviewertwocolor}
& $7.4 \pm 1.6\times 10^{-4}$ & $0.076$ &  $19.3$ &  $23.7$ & $0.81\times$ \\
& $m = 50000$
\rowstyle{\reviewertwocolor}
& $6.4 \pm 0.6\times 10^{-4}$ & $0.077$ &  $59.8$ &  $26.9$ & $2.22\times$ \\
& $m = 100000$
\rowstyle{\reviewertwocolor}
& $1.8 \pm 0.4\times 10^{-4}$ & $0.077$ &  $118.0$ &  $78.6$ & $1.50\times$ \\
& $m = 500000$
\rowstyle{\reviewertwocolor}
& $2.3 \pm 0.2\times 10^{-4}$ & $0.080$ &  $459.9$ &  $64.5$ & $7.13\times$ \\
& $m = 1000000$
\rowstyle{\reviewertwocolor}
& $5.4 \pm 1.2\times 10^{-5}$ & $0.083$ &  $872.4$ &  $196.5$ & $4.44\times$ \\
        \bottomrule
    \end{tabular}
    \caption{Convergence of ground-state energy for \ch{C2}.
    The reference ground-state energy is \(-75.7319615\) Ha.}
    \label{tab:c2}
\end{table*}


\begin{table*}[ht]
    \centering
    \begin{tabular}{=c =l|+r +r +r|+r +r}
        \toprule
        \multirow{2}{*}{\textbf{Algorithm}} & \multirow{2}{*}{\textbf{Parameter}}
        & \multirow{2}{*}{\textbf{Absolute Error}} & \multirow{2}{*}{\textbf{Mem(GB)}}
        & \multirow{2}{*}{\textbf{Time(s)}} & \multicolumn{2}{c}{\textbf{mCDFCI}} \\
        & & & & & \textbf{Time(s)} & \textbf{Ratio} \\
        \midrule
\multirow{5}{*}{mCDFCI}
& \multirow{5}{*}{$\tau = 5.0 \times 10^{-7}$}
  & $1.0 \times 10^{-2}$ & $3.0$ & $4.2$   & - & \\
& & $1.0 \times 10^{-3}$ & $12.0$ & $31.8$   & - & \\
& & $1.0 \times 10^{-4}$ & $24.0$ & $311.8$   & - & \\
& & $1.0 \times 10^{-5}$ & $24.0$ & $1830.3$   & - & \\
& & $1.0 \times 10^{-6}$ & $24.0$ & $5629.8$   & - & \\
\midrule
\multirow{5}{*}{sCDFCI}
& \multirow{5}{*}{$\varepsilon = 5.0 \times 10^{-7}$}
  & $1.0 \times 10^{-2}$ & $6.0$ & $19.2$ &  $4.2$ & $4.57\times$\\
& & $1.0 \times 10^{-3}$ & $12.0$ & $87.5$ &  $31.8$ & $2.75\times$\\
& & $1.0 \times 10^{-4}$ & $24.0$ & $829.7$ &  $311.8$ & $2.66\times$\\
& & $1.0 \times 10^{-5}$ & $24.0$ & $4687.7$ &  $1830.3$ & $2.56\times$\\
& & $1.0 \times 10^{-6}$ & $24.0$ & $14048.5$ &  $5629.8$ & $2.50\times$\\
\midrule
\multirow{5}{*}{SHCI}
& $\varepsilon_1 = 1.0\times 10^{-4}$
& $1.6\times 10^{-3}$ & $16.8$ & $4.0$ &  $19.7$ & $0.20\times$ \\
& $\varepsilon_1 = 5.0\times 10^{-5}$
& $8.8\times 10^{-4}$ & $18.9$ & $10.9$ &  $40.0$ & $0.27\times$ \\
& $\varepsilon_1 = 1.0\times 10^{-5}$
& $1.9\times 10^{-4}$ & $37.9$ & $76.2$ &  $175.2$ & $0.44\times$ \\
& $\varepsilon_1 = 5.0\times 10^{-6}$
& $8.6\times 10^{-5}$ & $68.2$ & $183.4$ &  $353.3$ & $0.52\times$ \\
& $\varepsilon_1 = 1.0\times 10^{-6}$
& $1.0\times 10^{-5}$ & $402.7$ & $1753.8$ &  $1796.5$ & $0.98\times$ \\
\midrule
\multirow{4}{*}{DMRG}
& $\max M = 500$
& $1.2\times 10^{-3}$ & $0.2$ & $126.5$ &  $27.9$ & $4.54\times$ \\
& $\max M = 1000$
& $3.1\times 10^{-4}$ & $0.8$ & $427.3$ &  $114.2$ & $3.74\times$ \\
& $\max M = 2000$
& $6.2\times 10^{-5}$ & $3.1$ & $1405.8$ &  $470.7$ & $2.99\times$ \\
& $\max M = 4000$
& $8.9\times 10^{-6}$ & $11.8$ & $4940.7$ &  $1976.8$ & $2.50\times$ \\
\midrule
\multirow{5}{*}{\reviewertwo{iS-FCIQMC}}
& $m = 10000$
\rowstyle{\reviewertwocolor}
& $4.4 \pm 1.4\times 10^{-4}$ & $0.162$ &  $31.3$ &  $79.6$ & $0.39\times$ \\
& $m = 50000$
\rowstyle{\reviewertwocolor}
& $2.4 \pm 0.5\times 10^{-4}$ & $0.162$ &  $108.9$ &  $141.0$ & $0.77\times$ \\
& $m = 100000$
\rowstyle{\reviewertwocolor}
& $1.2 \pm 0.3\times 10^{-4}$ & $0.163$ &  $233.0$ &  $270.1$ & $0.86\times$ \\
& $m = 500000$
\rowstyle{\reviewertwocolor}
& $7.5 \pm 1.4\times 10^{-5}$ & $0.166$ &  $858.7$ &  $398.8$ & $2.15\times$ \\
& $m = 1000000$
\rowstyle{\reviewertwocolor}
& $3.5 \pm 1.2\times 10^{-5}$ & $0.169$ &  $1596.3$ &  $735.0$ & $2.17\times$ \\
        \bottomrule
    \end{tabular}
    \caption{Convergence of ground-state energy for \ch{N2}.
    The reference ground-state energy is \(-109.2821721\) Ha.}
    \label{tab:n2}
\end{table*}


Table~\ref{tab:c2} and Table~\ref{tab:n2}
demonstrates the convergence behavior of the \ch{C2} and \ch{N2}
molecule respectively, executed by different methods.
Each method was evaluated with specific parameter values,
and the results are documented in terms of absolute error,
memory consumption in gigabytes (GB), execution time in seconds,
and a comparison with the mCDFCI algorithm regarding execution
time and relative ratio. From a broader perspective,
there are significant differences in time and memory usage
among the various algorithms, which we will examine in detail.

The proposed algorithm, mCDFCI, achieves chemical accuracy for the
\ch{C2} molecule in less than one minute while using only 6 GB of
memory. Subsequently, the energy continues to decrease at a linear rate.
The expansion of the primary variational space slows down
once the absolute error is less than \(10^{-5}\).
The wavefunction converges to a subspace of the
full CI space according to the current truncation threshold
\(\tau = 3 \times 10^{-8}\). For the \ch{N2} molecule,
mCDFCI exhibits similar behavior for threshold
\(\tau = 5 \times 10^{-7}\).

Compared to sCDFCI, which updates only one determinant per iteration,
mCDFCI descends much faster and utilizes a smaller subspace at the
start of the iterations. This demonstrates the efficiency of mCDFCI in
capturing important determinants. Towards the end
of the iterations, the effective iterations of mCDFCI and sCDFCI
are comparable. However, mCDFCI still shows approximately three
times the acceleration, highlighting the enhanced parallelization
power utilized by our approach.

SHCI also adopts the concept of selected CI but employs a different
data structure compared to ours.
After selecting determinants based on an importance measure
from the Hamiltonian matrix elements
connected to the reference determinant,
SHCI extracts a submatrix and directly diagonalizes it
to find the smallest eigenpair.
The efficiency of the SHCI algorithm can be
attributed to the powerful Davidson algorithm, the one-time
evaluation of the Hamiltonian matrix, and the contiguous memory
read/write operations. Additionally, multi-threading is particularly
effective for computing matrix-vector multiplications, making
the method highly parallel and further boosting the
performance. These factors enable SHCI to exhibit
speedups over mCDFCI, especially when the required accuracy is low.
However, the storage of the Hamiltonian submatrix leads to high
storage costs, particularly when high accuracy is required. On the
other hand, when \(\varepsilon_1 = 1\times 10^{5}\), SHCI uses
\(\num{3324630}\) determinants to reach an accuracy of
\(1.1 \times 10^{-4}\) Ha, whereas mCDFCI uses only \(\num{2022447}\)
determinants to achieve the same accuracy. In this case, mCDFCI
uses approximately \(40\%\) fewer determinants than SHCI to
attain the same level of precision.

DMRG, which models the wavefunction using a matrix product state form,
has a lower memory footprint than mCDFCI due to its storage cost of
\(\mathcal{O}(\norb M^2)\), where \(\norb\) is the number of orbitals
and \(M\) the maximum bond dimension.
However, mCDFCI achieves faster convergence compared to DMRG,
particularly when targeting moderate accuracy.
The difference in convergence speed decreases as higher accuracy
is required, but mCDFCI remains more efficient overall.

\reviewertwo{Finally, iS-FCIQMC, a stochastic method that
approximates the wavefunction using random walkers,
is highly memory-efficient, typically requiring a few megabytes.
While its error decreases gradually as the number of walkers increases,
iS-FCIQMC can achieve faster convergence for lower accuracy requirements.
On the other hand, mCDFCI, with its deterministic framework,
offers a reliable and efficient approach for achieving higher precision.}

\subsection{All-Electron Chromium Dimer Calculation}
We now turn our attention to the larger system of the chromium dimer,
which consists of \(48\) electrons and \(84\) spin--orbitals.
Transition metals play a pivotal role in catalysis, biochemistry,
and the energy sciences, but the complex electronic structure of
d-shells presents significant challenges for modeling and
understanding such processes~\cite{Larsson2022}.

In the case of the chromium dimer, the high electron correlation
arises from two main aspects: static correlation,
which is due to the spin coupling of the \(12\) valence electrons,
and dynamic correlation, which involves excitations of non-valence orbitals,
such as the 3p electrons. To address this challenge,
we computed the all-electron chromium dimer ground state
at a bond length of \(1.5\) \AA\ using the Ahlrichs SV basis set.

\begin{figure}[ht]
    \centering
    \includegraphics[width=0.5\textwidth]{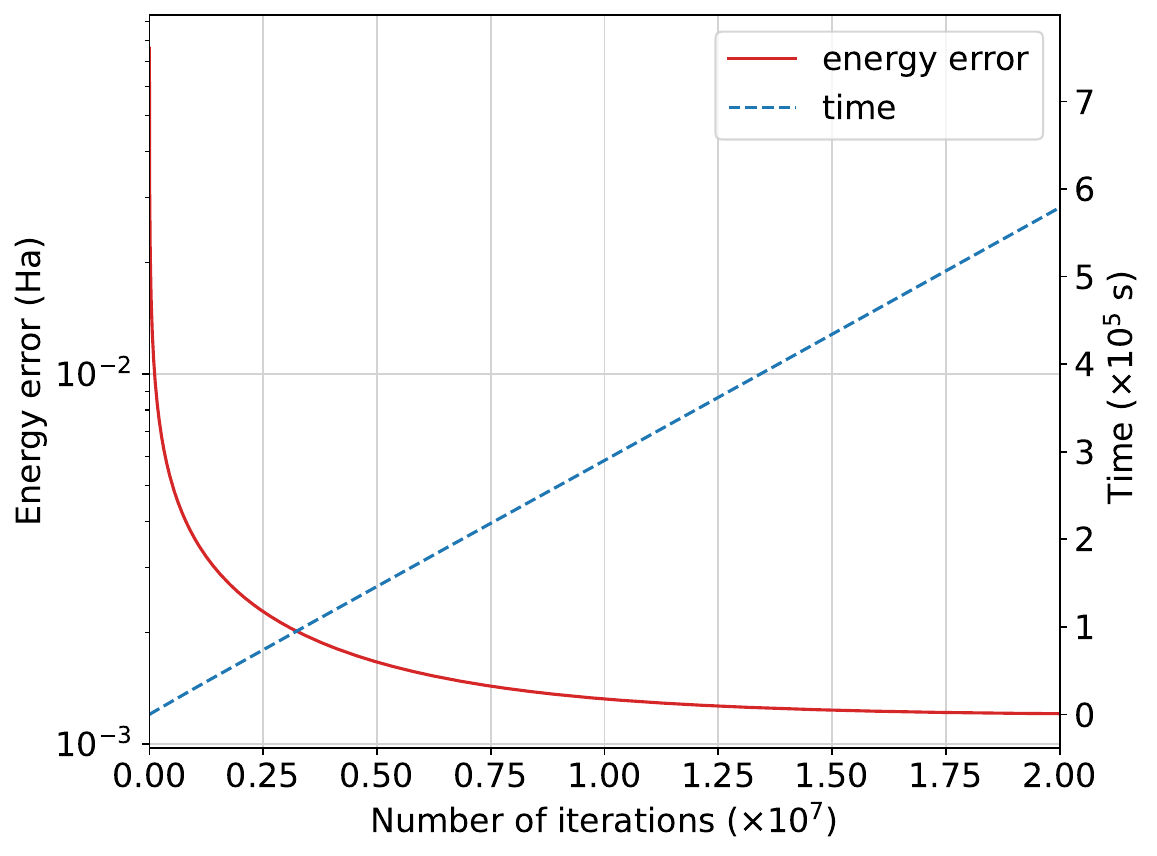}
    \caption{
        Convergence of ground-state energy for the \ch{Cr2}
        molecule under Ahlrichs SV basis. The threshold is set to
        \(3.7 \times 10^{-5}\) and the reference ground-state
        energy is \(-2086.444784\) Ha,
        obtained via extrapolated DMRG~\cite{Olivares-Amaya2015}.
    }
    \label{fig:cr2}
\end{figure}

Figure~\ref{fig:cr2} illustrates the energy decrease and time
usage for running \(2 \times 10^7\) iterations using \(128\)
threads. In this setup, \(128\) determinants are selected in each
iteration, for either addition to the variational space
or updating their associated values.
The final number of determinants in the variational wavefunction
is approximately \(2 \times 10^9\).

The red curve in the plot represents the absolute energy error (in Hartree)
as a function of the number of iterations.
\reviewerone{The energy error is computed as
the absolute difference between the current
variational energy (Rayleigh quotient)
and the reference ground-state energy.}
This error decreases rapidly to the level of \(10^{-3}\) within several hours,
before converging towards the ground-state energy at a steady linear rate.

The blue dashed curve in the plot shows the total time elapsed as a function
of the number of iterations, measured in seconds (\(\times 10^5\)).
A total of \(2 \times 10^7\) iterations takes around \(6 \times 10^5\)
seconds, which is roughly one week.
The linear trend indicates that the time per iteration remains
relatively constant.
This consistency is due to the number of nonzeros in each column of
the Hamiltonian matrix being similar (around \(3 \times 10^4\)),
leading to a nearly constant computational cost per iteration
(See Table~\ref{tab:complexity} for complexity analysis).
This predictable linear increase in time per iteration allows users
to estimate the remaining computation time easily,
demonstrating the practicality of the CDFCI algorithm.

Compared to the previous CDFCI algorithm with single-coordinate updates,
which achieved an energy of \(-2086.443565\) Ha
after one month of computation~\cite{Wang2019}, we reached
the same level of accuracy in just \(5.8\) days,
despite using different computing environments.
Additionally, we obtained a more accurate energy of
\(-2086.443581\) Ha in approximately \(8.8\) days.

\subsection{Binding Curve of \ch{N2} under cc-pVQZ Basis}

Finally, we benchmark the all-electron nitrogen binding curve
under cc-pVQZ basis.
The problem is challenging due to the complexity of accurately
capturing the strong electron correlation in the triple bond of \ch{N2}.
The large cc-pVQZ basis set increases computational cost,
and all-electron calculations add further complexity by
requiring the explicit treatment of core electrons.
We adopt \(\tau = 5 \times 10^{-6}\) for mCDFCI truncation,
use \(128\) threads, and run \(\num{1000000}\) iterations.
The bond length is varied from \(1.5\) to \(4.5 \, a_0\).
Each configuration on the binding curve takes roughly
\(18\) hours to achieve chemical accuracy.

\begin{figure}[ht]
    \centering
    \includegraphics[width=0.5\textwidth]{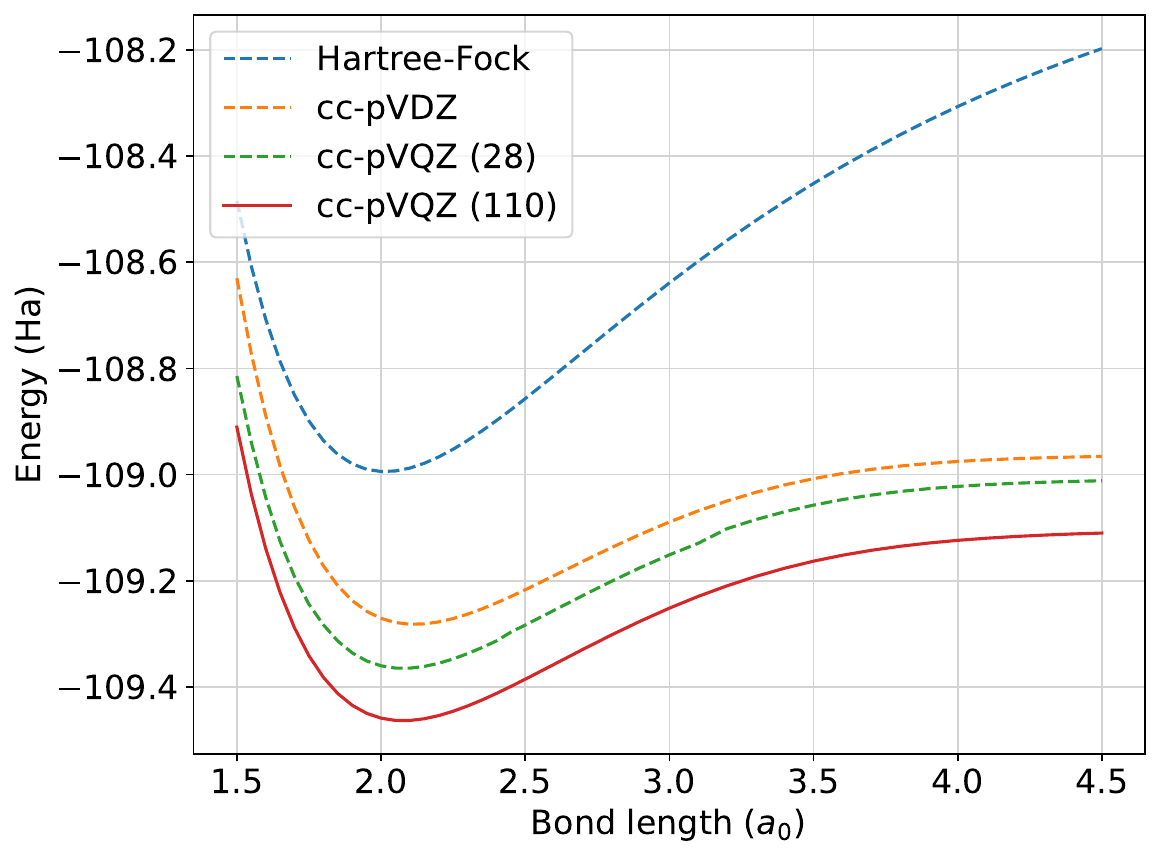}
    \caption{
        ground-state energy versus bond length for \ch{N2}
        using different basis sets.
        cc-pVQZ (110) refers to the full,
        uncompressed cc-pVQZ basis set, while cc-pVQZ (28)
        refers to the compressed set of orbitals
        generated by OptOrbFCI.
        The mCDFCI truncation threshold
        \(\tau = 5 \times 10^{-6}\) and stopping criteria
        of \(1 \times 10^{-5}\) are applied.
    }
    \label{fig:dissociation}
\end{figure}

Figure~\ref{fig:dissociation}
displays a smooth, standard-shaped binding curve (shown in red)
for \ch{N2} calculated using the mCDFCI method
with a full cc-pVQZ basis set (110 orbitals).
The curve reaches its minimum at the equilibrium
bond length of \(r = 2.118 \, a_0\), which corresponds to
the optimal bond length where the energy is minimized.
This red curve is compared against other computational results,
including the Hartree--Fock approximation (blue dashed line)
and CDFCI calculations using a reduced orbital set generated from cc-pVQZ,
which is compressed to 28 selected orbitals (green dashed curve).

The Hartree--Fock method overestimates the total energy,
particularly at longer bond lengths,
due to its inability to capture dynamic electron correlation
and its inherent inclusion of ionic character~\cite{Kreplin2020}.
The green dashed curve represents results from the
OptOrbFCI method~\cite{Li2020}, which uses a rotation and compression
of the full cc-pVQZ basis set into 28 selected orbitals.
This method was employed to reduce the computational cost while
retaining the essential characteristics of the electronic structure
of the system. Despite the reduction in the number of orbitals,
OptOrbFCI produces an energy profile closer to the
full cc-pVQZ  calculation than the cc-pVDZ approximation,
showing that the selected orbitals capture most of the correlation effects.
However, the slight increase in energy compared to the full
cc-pVQZ curve reflects the inherent trade-off between
computational efficiency and accuracy
when compressing the basis set.

Appendix~\ref{app:dissociation} lists all the variational energies
corresponding to each bond length and configuration shown in the figure,
providing a detailed comparison of the methods and basis sets.
This comparison underscores the importance of increasing the basis set
size to achieve basis set convergence in quantum chemical calculations.

\section{Conclusion}
\label{sec:conclusion}

This paper presents a novel algorithm,
multi-Coordinate Descent FCI method,
for solving the full configuration interaction (FCI) problem
in electronic structure calculations.
The algorithm reformulates the FCI eigenvalue problem
as an unconstrained optimization problem and
employs a multi-coordinate descent approach
to update the optimization variable. Additionally, the algorithm
introduces a compression strategy to control
the size of the variational space.

The method demonstrates efficiency and accuracy through various
benchmarks, including the computation of an accurate
benchmark energy for the chromium dimer and the binding
curve of nitrogen dimer. The results show up to \(79.3\%\)
parallel efficiency on \(128\) cores, as well as significant speedup
compared to previous methods. Overall, the Multi-Coordinate Descent
FCI method presents a promising approach for solving the FCI problem
in electronic structure calculations, with improved efficiency, accuracy,
and parallelization capability, making it suitable for larger systems
and more accurate approximations.

Since the current algorithm focuses on improving the
variational stage of selected CI,
our immediate future work will involve adding a perturbation stage
to the current algorithm. Another direction will be to combine the
current results with orbital optimization,
as it is well known that optimized orbitals lead to faster
convergence.~\cite{Smith2017, JLi2020, Li2020}.
Several approaches for computing excited states within the FCI framework have
been developed~\cite{Blunt2015, Holmes2017, Schriber2017, Gao2022, Gao2024}.
These methods, often extensions of ground-state algorithms,
have demonstrated success in accurately treating excited states.
Recently, Wang et al.\ introduced xCDFCI~\cite{Wang2023},
an efficient extension of CDFCI designed specifically
for low-lying excited states in molecules.
We aim to adapt our techniques for excited-state computations as well.

Regarding parallel capacity, extending our current shared-memory
implementation to a distributed-memory one with MPI enabled is
an interesting prospect, as it would further increase computational
capacity and leverage memory load. A well-implemented distributed
memory setup is under investigation.

\section*{Acknowledgments}

The authors thank Jianfeng Lu and Zhe Wang for their helpful discussions
and valuable insights during the course of this work.
This research was supported by
the National Natural Science Foundation of China (NSFC)
under grant numbers 12271109 and 71991471,
the National Key R\&D Program of China under grant 2020YFA0711902,
the Science and Technology Commission of Shanghai Municipality (STCSM)
under grant numbers 22TQ017 and 24DP2600100, and the
Shanghai Institute for Mathematics and Interdisciplinary Sciences (SIMIS)
under grant number SIMIS-ID-2024-(CN).
The authors are also grateful for the resources and facilities
provided by NSFC, the National Key R\&D Program of China, STCSM, and SIMIS,
which were essential for the completion of this work.

\bibliography{main}

\providecommand{\latin}[1]{#1}
\makeatletter
\providecommand{\doi}
  {\begingroup\let\do\@makeother\dospecials
  \catcode`\{=1 \catcode`\}=2 \doi@aux}
\providecommand{\doi@aux}[1]{\endgroup\texttt{#1}}
\makeatother
\providecommand*\mcitethebibliography{\thebibliography}
\csname @ifundefined\endcsname{endmcitethebibliography}
  {\let\endmcitethebibliography\endthebibliography}{}
\begin{mcitethebibliography}{40}
\providecommand*\natexlab[1]{#1}
\providecommand*\mciteSetBstSublistMode[1]{}
\providecommand*\mciteSetBstMaxWidthForm[2]{}
\providecommand*\mciteBstWouldAddEndPuncttrue
  {\def\EndOfBibitem{\unskip.}}
\providecommand*\mciteBstWouldAddEndPunctfalse
  {\let\EndOfBibitem\relax}
\providecommand*\mciteSetBstMidEndSepPunct[3]{}
\providecommand*\mciteSetBstSublistLabelBeginEnd[3]{}
\providecommand*\EndOfBibitem{}
\mciteSetBstSublistMode{f}
\mciteSetBstMaxWidthForm{subitem}{(\alph{mcitesubitemcount})}
\mciteSetBstSublistLabelBeginEnd
  {\mcitemaxwidthsubitemform\space}
  {\relax}
  {\relax}

\bibitem[White and Martin(1999)White, and Martin]{White1999}
White,~S.~R.; Martin,~R.~L. Ab initio quantum chemistry using the density
  matrix renormalization group. \emph{The Journal of Chemical Physics}
  \textbf{1999}, \emph{110}, 4127--4130\relax
\mciteBstWouldAddEndPuncttrue
\mciteSetBstMidEndSepPunct{\mcitedefaultmidpunct}
{\mcitedefaultendpunct}{\mcitedefaultseppunct}\relax
\EndOfBibitem
\bibitem[Chan and Head-Gordon(2002)Chan, and Head-Gordon]{Chan2002}
Chan,~G. K.-L.; Head-Gordon,~M. Highly correlated calculations with a
  polynomial cost algorithm: A study of the density matrix renormalization
  group. \emph{The Journal of Chemical Physics} \textbf{2002}, \emph{116},
  4462--4476\relax
\mciteBstWouldAddEndPuncttrue
\mciteSetBstMidEndSepPunct{\mcitedefaultmidpunct}
{\mcitedefaultendpunct}{\mcitedefaultseppunct}\relax
\EndOfBibitem
\bibitem[Chan and Sharma(2011)Chan, and Sharma]{Chan2011}
Chan,~G. K.-L.; Sharma,~S. The Density Matrix Renormalization Group in Quantum
  Chemistry. \emph{Annual Review of Physical Chemistry} \textbf{2011},
  \emph{62}, 465--481\relax
\mciteBstWouldAddEndPuncttrue
\mciteSetBstMidEndSepPunct{\mcitedefaultmidpunct}
{\mcitedefaultendpunct}{\mcitedefaultseppunct}\relax
\EndOfBibitem
\bibitem[Schollwöck(2011)]{Schollwock2011}
Schollwöck,~U. The density-matrix renormalization group in the age of matrix
  product states. \emph{Annals of Physics} \textbf{2011}, \emph{326},
  96--192\relax
\mciteBstWouldAddEndPuncttrue
\mciteSetBstMidEndSepPunct{\mcitedefaultmidpunct}
{\mcitedefaultendpunct}{\mcitedefaultseppunct}\relax
\EndOfBibitem
\bibitem[Keller \latin{et~al.}(2015)Keller, Dolfi, Troyer, and
  Reiher]{Keller2015}
Keller,~S.; Dolfi,~M.; Troyer,~M.; Reiher,~M. An efficient matrix product
  operator representation of the quantum chemical Hamiltonian. \emph{The
  Journal of Chemical Physics} \textbf{2015}, \emph{143}, 244118\relax
\mciteBstWouldAddEndPuncttrue
\mciteSetBstMidEndSepPunct{\mcitedefaultmidpunct}
{\mcitedefaultendpunct}{\mcitedefaultseppunct}\relax
\EndOfBibitem
\bibitem[Booth \latin{et~al.}(2009)Booth, Thom, and Alavi]{Booth2009}
Booth,~G.~H.; Thom,~A. J.~W.; Alavi,~A. Fermion Monte Carlo without fixed
  nodes: A game of life, death, and annihilation in Slater determinant space.
  \emph{The Journal of Chemical Physics} \textbf{2009}, \emph{131},
  054106\relax
\mciteBstWouldAddEndPuncttrue
\mciteSetBstMidEndSepPunct{\mcitedefaultmidpunct}
{\mcitedefaultendpunct}{\mcitedefaultseppunct}\relax
\EndOfBibitem
\bibitem[Ghanem \latin{et~al.}(2019)Ghanem, Lozovoi, and Alavi]{Ghanem2019}
Ghanem,~K.; Lozovoi,~A.~Y.; Alavi,~A. Unbiasing the initiator approximation in
  full configuration interaction quantum Monte Carlo. \emph{The Journal of
  Chemical Physics} \textbf{2019}, \emph{151}, 224108\relax
\mciteBstWouldAddEndPuncttrue
\mciteSetBstMidEndSepPunct{\mcitedefaultmidpunct}
{\mcitedefaultendpunct}{\mcitedefaultseppunct}\relax
\EndOfBibitem
\bibitem[Blunt \latin{et~al.}(2019)Blunt, Thom, and Scott]{Blunt2019}
Blunt,~N.~S.; Thom,~A. J.~W.; Scott,~C. J.~C. Preconditioning and Perturbative
  Estimators in Full Configuration Interaction Quantum Monte Carlo.
  \emph{Journal of Chemical Theory and Computation} \textbf{2019}, \emph{15},
  3537--3551\relax
\mciteBstWouldAddEndPuncttrue
\mciteSetBstMidEndSepPunct{\mcitedefaultmidpunct}
{\mcitedefaultendpunct}{\mcitedefaultseppunct}\relax
\EndOfBibitem
\bibitem[Cleland \latin{et~al.}(2010)Cleland, Booth, and Alavi]{Cleland2010}
Cleland,~D.; Booth,~G.~H.; Alavi,~A. Communications: Survival of the fittest:
  Accelerating convergence in full configuration-interaction quantum Monte
  Carlo. \emph{The Journal of Chemical Physics} \textbf{2010}, \emph{132},
  041103\relax
\mciteBstWouldAddEndPuncttrue
\mciteSetBstMidEndSepPunct{\mcitedefaultmidpunct}
{\mcitedefaultendpunct}{\mcitedefaultseppunct}\relax
\EndOfBibitem
\bibitem[Yang \latin{et~al.}(2020)Yang, Pahl, and Brand]{Yang2020}
Yang,~M.; Pahl,~E.; Brand,~J. Improved walker population control for full
  configuration interaction quantum Monte Carlo. \emph{The Journal of Chemical
  Physics} \textbf{2020}, \emph{153}, 174103\relax
\mciteBstWouldAddEndPuncttrue
\mciteSetBstMidEndSepPunct{\mcitedefaultmidpunct}
{\mcitedefaultendpunct}{\mcitedefaultseppunct}\relax
\EndOfBibitem
\bibitem[Eriksen(2021)]{Eriksen2021}
Eriksen,~J.~J. The Shape of Full Configuration Interaction to Come. \emph{The
  Journal of Physical Chemistry Letters} \textbf{2021}, \emph{12},
  418--432\relax
\mciteBstWouldAddEndPuncttrue
\mciteSetBstMidEndSepPunct{\mcitedefaultmidpunct}
{\mcitedefaultendpunct}{\mcitedefaultseppunct}\relax
\EndOfBibitem
\bibitem[Huron \latin{et~al.}(1973)Huron, Malrieu, and Rancurel]{Huron1973}
Huron,~B.; Malrieu,~J.; Rancurel,~P. Iterative perturbation calculations of
  ground and excited state energies from multiconfigurational zeroth-order
  wavefunctions. \emph{The Journal of Chemical Physics} \textbf{1973},
  \emph{58}, 5745--5759\relax
\mciteBstWouldAddEndPuncttrue
\mciteSetBstMidEndSepPunct{\mcitedefaultmidpunct}
{\mcitedefaultendpunct}{\mcitedefaultseppunct}\relax
\EndOfBibitem
\bibitem[Tubman \latin{et~al.}(2016)Tubman, Lee, Takeshita, Head-Gordon, and
  Whaley]{Tubman2016}
Tubman,~N.~M.; Lee,~J.; Takeshita,~T.~Y.; Head-Gordon,~M.; Whaley,~K.~B. A
  deterministic alternative to the full configuration interaction quantum Monte
  Carlo method. \emph{The Journal of Chemical Physics} \textbf{2016},
  \emph{145}\relax
\mciteBstWouldAddEndPuncttrue
\mciteSetBstMidEndSepPunct{\mcitedefaultmidpunct}
{\mcitedefaultendpunct}{\mcitedefaultseppunct}\relax
\EndOfBibitem
\bibitem[Holmes \latin{et~al.}(2016)Holmes, Tubman, and Umrigar]{Holmes2016}
Holmes,~A.~A.; Tubman,~N.~M.; Umrigar,~C. Heat-bath configuration interaction:
  An efficient selected configuration interaction algorithm inspired by
  heat-bath sampling. \emph{Journal of Chemical Theory and Computation}
  \textbf{2016}, \emph{12}, 3674--3680\relax
\mciteBstWouldAddEndPuncttrue
\mciteSetBstMidEndSepPunct{\mcitedefaultmidpunct}
{\mcitedefaultendpunct}{\mcitedefaultseppunct}\relax
\EndOfBibitem
\bibitem[Sharma \latin{et~al.}(2017)Sharma, Holmes, Jeanmairet, Alavi, and
  Umrigar]{Sharma2017}
Sharma,~S.; Holmes,~A.~A.; Jeanmairet,~G.; Alavi,~A.; Umrigar,~C.~J.
  Semistochastic heat-bath configuration interaction method: Selected
  configuration interaction with semistochastic perturbation theory.
  \emph{Journal of Chemical Theory and Computation} \textbf{2017}, \emph{13},
  1595--1604\relax
\mciteBstWouldAddEndPuncttrue
\mciteSetBstMidEndSepPunct{\mcitedefaultmidpunct}
{\mcitedefaultendpunct}{\mcitedefaultseppunct}\relax
\EndOfBibitem
\bibitem[Li \latin{et~al.}(2018)Li, Otten, Holmes, Sharma, and Umrigar]{Li2018}
Li,~J.; Otten,~M.; Holmes,~A.~A.; Sharma,~S.; Umrigar,~C.~J. Fast
  Semistochastic Heat-Bath Configuration Interaction. \emph{The Journal of
  Chemical Physics} \textbf{2018}, \emph{148}, 214110\relax
\mciteBstWouldAddEndPuncttrue
\mciteSetBstMidEndSepPunct{\mcitedefaultmidpunct}
{\mcitedefaultendpunct}{\mcitedefaultseppunct}\relax
\EndOfBibitem
\bibitem[Wang \latin{et~al.}(2019)Wang, Li, and Lu]{Wang2019}
Wang,~Z.; Li,~Y.; Lu,~J. Coordinate Descent Full Configuration Interaction.
  \emph{Journal of Chemical Theory and Computation} \textbf{2019}, \emph{15},
  3558--3569\relax
\mciteBstWouldAddEndPuncttrue
\mciteSetBstMidEndSepPunct{\mcitedefaultmidpunct}
{\mcitedefaultendpunct}{\mcitedefaultseppunct}\relax
\EndOfBibitem
\bibitem[Greene \latin{et~al.}(2020)Greene, Webber, Weare, and
  Berkelbach]{Greene2020}
Greene,~S.~M.; Webber,~R.~J.; Weare,~J.; Berkelbach,~T.~C. Improved Fast
  Randomized Iteration Approach to Full Configuration Interaction.
  \emph{Journal of Chemical Theory and Computation} \textbf{2020}, \emph{16},
  5572--5585\relax
\mciteBstWouldAddEndPuncttrue
\mciteSetBstMidEndSepPunct{\mcitedefaultmidpunct}
{\mcitedefaultendpunct}{\mcitedefaultseppunct}\relax
\EndOfBibitem
\bibitem[Goings \latin{et~al.}(2021)Goings, Hu, Yang, and Li]{Goings2021}
Goings,~J.~J.; Hu,~H.; Yang,~C.; Li,~X. Reinforcement Learning Configuration
  Interaction. \emph{Journal of Chemical Theory and Computation} \textbf{2021},
  \emph{17}, 5482--5491\relax
\mciteBstWouldAddEndPuncttrue
\mciteSetBstMidEndSepPunct{\mcitedefaultmidpunct}
{\mcitedefaultendpunct}{\mcitedefaultseppunct}\relax
\EndOfBibitem
\bibitem[Li \latin{et~al.}(2019)Li, Lu, and Wang]{Li2019}
Li,~Y.; Lu,~J.; Wang,~Z. Coordinate-Wise Descent Methods for Leading Eigenvalue
  Problem. \emph{SIAM Journal on Scientific Computing} \textbf{2019},
  \emph{41}, A2681--A2716\relax
\mciteBstWouldAddEndPuncttrue
\mciteSetBstMidEndSepPunct{\mcitedefaultmidpunct}
{\mcitedefaultendpunct}{\mcitedefaultseppunct}\relax
\EndOfBibitem
\bibitem[Dhillon \latin{et~al.}(2006)Dhillon, Parlett, and V{\"{o}}mel]{MRRR}
Dhillon,~I.~S.; Parlett,~B.~N.; V{\"{o}}mel,~C. The design and implementation
  of the MRRR algorithm. \emph{ACM Transactions on Mathematical Software}
  \textbf{2006}, \emph{32}, 533--560\relax
\mciteBstWouldAddEndPuncttrue
\mciteSetBstMidEndSepPunct{\mcitedefaultmidpunct}
{\mcitedefaultendpunct}{\mcitedefaultseppunct}\relax
\EndOfBibitem
\bibitem[Li and Lu(2020)Li, and Lu]{Li2020}
Li,~Y.; Lu,~J. Optimal Orbital Selection for Full Configuration Interaction
  (OptOrbFCI): Pursuing the Basis Set Limit under a Budget. \emph{Journal of
  Chemical Theory and Computation} \textbf{2020}, \emph{16}, 6207--6221\relax
\mciteBstWouldAddEndPuncttrue
\mciteSetBstMidEndSepPunct{\mcitedefaultmidpunct}
{\mcitedefaultendpunct}{\mcitedefaultseppunct}\relax
\EndOfBibitem
\bibitem[lib()]{libcuckoo}
A high-performance, concurrent hash table.
  \url{https://github.com/efficient/libcuckoo}, Accessed: 2024-05-07\relax
\mciteBstWouldAddEndPuncttrue
\mciteSetBstMidEndSepPunct{\mcitedefaultmidpunct}
{\mcitedefaultendpunct}{\mcitedefaultseppunct}\relax
\EndOfBibitem
\bibitem[Turney \latin{et~al.}(2012)Turney, Simmonett, Parrish, Hohenstein,
  Evangelista, Fermann, Mintz, Burns, Wilke, Abrams, Russ, Leininger, Janssen,
  Seidl, Allen, Schaefer, King, Valeev, Sherrill, and Crawford]{psi4}
Turney,~J.~M. \latin{et~al.}  Psi4: an open-source ab initio electronic
  structure program. \emph{WIREs Computational Molecular Science}
  \textbf{2012}, \emph{2}, 556--565\relax
\mciteBstWouldAddEndPuncttrue
\mciteSetBstMidEndSepPunct{\mcitedefaultmidpunct}
{\mcitedefaultendpunct}{\mcitedefaultseppunct}\relax
\EndOfBibitem
\bibitem[cdf()]{cdfci}
C++ implemented Coordinate Descent Full Configuration Interaction (CDFCI)
  package for quantum chemistry calculation.
  \url{https://github.com/quan-tum/CDFCI}, Accessed: 2024-06-21\relax
\mciteBstWouldAddEndPuncttrue
\mciteSetBstMidEndSepPunct{\mcitedefaultmidpunct}
{\mcitedefaultendpunct}{\mcitedefaultseppunct}\relax
\EndOfBibitem
\bibitem[shc()]{shci}
Arrow - Fast Semistochastic Heat Bath Configuration Interaction Solver (SHCI).
  \url{https://github.com/QMC-Cornell/shci}, Accessed: 2024-06-21\relax
\mciteBstWouldAddEndPuncttrue
\mciteSetBstMidEndSepPunct{\mcitedefaultmidpunct}
{\mcitedefaultendpunct}{\mcitedefaultseppunct}\relax
\EndOfBibitem
\bibitem[dmr()]{dmrg}
Efficient parallel quantum chemistry DMRG in MPO formalism.
  \url{https://github.com/block-hczhai/block2-preview}, Accessed:
  2024-06-21\relax
\mciteBstWouldAddEndPuncttrue
\mciteSetBstMidEndSepPunct{\mcitedefaultmidpunct}
{\mcitedefaultendpunct}{\mcitedefaultseppunct}\relax
\EndOfBibitem
\bibitem[fci()]{fciqmc}
Standalone NECI codebase designed for FCIQMC and other stochastic quantum
  chemistry methods. \url{https://github.com/quan-tum/CDFCI}, Accessed:
  2024-06-21\relax
\mciteBstWouldAddEndPuncttrue
\mciteSetBstMidEndSepPunct{\mcitedefaultmidpunct}
{\mcitedefaultendpunct}{\mcitedefaultseppunct}\relax
\EndOfBibitem
\bibitem[Larsson \latin{et~al.}(2022)Larsson, Zhai, Umrigar, and
  Chan]{Larsson2022}
Larsson,~H.~R.; Zhai,~H.; Umrigar,~C.~J.; Chan,~G. K.-L. The Chromium Dimer:
  Closing a Chapter of Quantum Chemistry. \emph{Journal of the American
  Chemical Society} \textbf{2022}, \emph{144}, 15932--15937\relax
\mciteBstWouldAddEndPuncttrue
\mciteSetBstMidEndSepPunct{\mcitedefaultmidpunct}
{\mcitedefaultendpunct}{\mcitedefaultseppunct}\relax
\EndOfBibitem
\bibitem[Olivares-Amaya \latin{et~al.}(2015)Olivares-Amaya, Hu, Nakatani,
  Sharma, Yang, and Chan]{Olivares-Amaya2015}
Olivares-Amaya,~R.; Hu,~W.; Nakatani,~N.; Sharma,~S.; Yang,~J.; Chan,~G. K.-L.
  The ab-initio density matrix renormalization group in practice. \emph{The
  Journal of Chemical Physics} \textbf{2015}, \emph{142}, 034102\relax
\mciteBstWouldAddEndPuncttrue
\mciteSetBstMidEndSepPunct{\mcitedefaultmidpunct}
{\mcitedefaultendpunct}{\mcitedefaultseppunct}\relax
\EndOfBibitem
\bibitem[Kreplin(2020)]{Kreplin2020}
Kreplin,~D.~A. Multiconfiguration Self-consistent Field Methods for Large
  Molecules. Ph.D.\ thesis, Universit{\"a}t Stuttgart, 2020\relax
\mciteBstWouldAddEndPuncttrue
\mciteSetBstMidEndSepPunct{\mcitedefaultmidpunct}
{\mcitedefaultendpunct}{\mcitedefaultseppunct}\relax
\EndOfBibitem
\bibitem[Smith \latin{et~al.}(2017)Smith, Mussard, Holmes, and
  Sharma]{Smith2017}
Smith,~J. E.~T.; Mussard,~B.; Holmes,~A.~A.; Sharma,~S. Cheap and Near Exact
  CASSCF with Large Active Spaces. \emph{Journal of Chemical Theory and
  Computation} \textbf{2017}, \emph{13}, 5468--5478\relax
\mciteBstWouldAddEndPuncttrue
\mciteSetBstMidEndSepPunct{\mcitedefaultmidpunct}
{\mcitedefaultendpunct}{\mcitedefaultseppunct}\relax
\EndOfBibitem
\bibitem[Li \latin{et~al.}(2020)Li, Yao, Holmes, Otten, Sun, Sharma, and
  Umrigar]{JLi2020}
Li,~J.; Yao,~Y.; Holmes,~A.~A.; Otten,~M.; Sun,~Q.; Sharma,~S.; Umrigar,~C.~J.
  Accurate many-body electronic structure near the basis set limit: Application
  to the chromium dimer. \emph{Physical Review Research} \textbf{2020},
  \emph{2}, 012015\relax
\mciteBstWouldAddEndPuncttrue
\mciteSetBstMidEndSepPunct{\mcitedefaultmidpunct}
{\mcitedefaultendpunct}{\mcitedefaultseppunct}\relax
\EndOfBibitem
\bibitem[Blunt \latin{et~al.}(2015)Blunt, Smart, Booth, and Alavi]{Blunt2015}
Blunt,~N.~S.; Smart,~S.~D.; Booth,~G.~H.; Alavi,~A. An excited-state approach
  within full configuration interaction quantum Monte Carlo. \emph{The Journal
  of Chemical Physics} \textbf{2015}, \emph{143}, 134117\relax
\mciteBstWouldAddEndPuncttrue
\mciteSetBstMidEndSepPunct{\mcitedefaultmidpunct}
{\mcitedefaultendpunct}{\mcitedefaultseppunct}\relax
\EndOfBibitem
\bibitem[Holmes \latin{et~al.}(2017)Holmes, Umrigar, and Sharma]{Holmes2017}
Holmes,~A.~A.; Umrigar,~C.~J.; Sharma,~S. Excited states using semistochastic
  heat-bath configuration interaction. \emph{The Journal of Chemical Physics}
  \textbf{2017}, \emph{147}, 164111\relax
\mciteBstWouldAddEndPuncttrue
\mciteSetBstMidEndSepPunct{\mcitedefaultmidpunct}
{\mcitedefaultendpunct}{\mcitedefaultseppunct}\relax
\EndOfBibitem
\bibitem[Schriber and Evangelista(2017)Schriber, and Evangelista]{Schriber2017}
Schriber,~J.~B.; Evangelista,~F.~A. Adaptive configuration interaction for
  computing challenging electronic excited states with tunable accuracy.
  \emph{Journal of Chemical Theory and Computation} \textbf{2017}, \emph{13},
  5354--5366\relax
\mciteBstWouldAddEndPuncttrue
\mciteSetBstMidEndSepPunct{\mcitedefaultmidpunct}
{\mcitedefaultendpunct}{\mcitedefaultseppunct}\relax
\EndOfBibitem
\bibitem[Gao \latin{et~al.}(2022)Gao, Li, and Lu]{Gao2022}
Gao,~W.; Li,~Y.; Lu,~B. Triangularized orthogonalization-free method for
  solving extreme eigenvalue problems. \emph{Journal of Scientific Computing}
  \textbf{2022}, \emph{93}\relax
\mciteBstWouldAddEndPuncttrue
\mciteSetBstMidEndSepPunct{\mcitedefaultmidpunct}
{\mcitedefaultendpunct}{\mcitedefaultseppunct}\relax
\EndOfBibitem
\bibitem[Gao \latin{et~al.}(2024)Gao, Li, and Shen]{Gao2024}
Gao,~W.; Li,~Y.; Shen,~H. Weighted trace-penalty minimization for full
  configuration interaction. \emph{SIAM Journal on Scientific Computing}
  \textbf{2024}, \emph{46}, A179--A203\relax
\mciteBstWouldAddEndPuncttrue
\mciteSetBstMidEndSepPunct{\mcitedefaultmidpunct}
{\mcitedefaultendpunct}{\mcitedefaultseppunct}\relax
\EndOfBibitem
\bibitem[Wang \latin{et~al.}(2023)Wang, Zhang, Lu, and Li]{Wang2023}
Wang,~Z.; Zhang,~Z.; Lu,~J.; Li,~Y. Coordinate Descent Full Configuration
  Interaction for Excited States. \emph{Journal of Chemical Theory and
  Computation} \textbf{2023}, \emph{19}, 7731--7739\relax
\mciteBstWouldAddEndPuncttrue
\mciteSetBstMidEndSepPunct{\mcitedefaultmidpunct}
{\mcitedefaultendpunct}{\mcitedefaultseppunct}\relax
\EndOfBibitem
\end{mcitethebibliography}

\newpage
\appendix

\section{Proof in Section~\ref{sec:coord-update}}
\label{app:proof}
We prove \(\lambda_{\min}((\at{Q}{\itellp})^\T H \at{Q}{\itellp}) < 0\)
as long as we choose an initial vector \(\Bat{c}{1}\) such that
\((\Bat{c}{1})^\T H \Bat{c}{1} < 0\), where \(\at{Q}{\itellp}\)
is defined in~\eqref{eq:define-q}.
The proof uses induction and contradiction.
In simple terms, if we assume \((\at{Q}{\itellp})^\T H \at{Q}{\itellp}\)
is positive semi-definite, then the minimization problem
will attain its minimum when the optimization vector is zero.
However, we can always find a vector that leads to a smaller
value, thus contradicting our initial assumption.

We first show that if \(\Bat{c}{1}\) is chosen such that
\((\Bat{c}{1})^\T H \Bat{c}{1} < 0\), then,
starting from the first iteration, we have
\(\lambda_{\min}((\at{Q}{1})^\T H \at{Q}{1}) < 0\).
Conversely, if all eigenvalues of
\((\at{Q}{1})^\T H \at{Q}{1}\) are non-negative, solving the
minimization problem~\eqref{eq:optimization-z} yields
\(\Bat{z}{1} = \B{0}\) and \(f(\Bat{z}{1}) = \|H\|^2_\F\).
However, \(\|H\|^2_\F\) is clearly not the minimal
point for \(f(\Bat{c}{2})\). To see this, we take
\[
    \at{\gamma}{1} = \frac{\sqrt{-(\Bat{c}{1})^\T H \Bat{c}{1}}}
        {(\Bat{c}{1})^\T \Bat{c}{1}},
\]
and \(\Bat{a}{1} = \B{0}\), we have
\[
    \begin{aligned}
    f(\Bat{c}{2}) &= \left\|H+\at{\gamma}{1} \Bat{c}{1}
    (\at{\gamma}{1} \Bat{c}{1})^\T\right\|^2_\F \\
    &= \|H\|^2_\F + 2 (\at{\gamma}{1})^2
    (\Bat{c}{1})^\T H \Bat{c}{1} \\
    &\quad + (\at{\gamma}{1})^4
    ((\Bat{c}{1})^\T \Bat{c}{1})^2 \\
    &= \|H\|^2_\F - \left(\frac{(\Bat{c}{1})^\T
    H \Bat{c}{1}}{(\Bat{c}{1})^\T \Bat{c}{1}}\right)^2 \\
    & < \|H\|^2_\F.
    \end{aligned}
\]
Thus \((\at{Q}{1})^\T H \at{Q}{1}\)  must have at least one negative
eigenvalue.

We then show that if, in the previous step, we have
\(\lambda_{\min}((\at{Q}{\itell})^\T H \at{Q}{\itell}) < 0\),
it follows that
\(\lambda_{\min}((\at{Q}{\itellp})^\T H \at{Q}{\itellp}) < 0\).
The reasoning is similar to above. Suppose this is not true,
such that all eigenvalues of \((\at{Q}{\itellp})^\T H \at{Q}{\itellp}\)
are non-negative.
Solving the minimization problem~\eqref{eq:optimization-z} would result in
\(\Bat{z}{\itellp} = \B{0}\) and \(f(\Bat{z}{(\itellp)}) = \|H\|^2_\F\).
However, this is impossible. To see why, consider taking
\(\at{\gamma}{\itellp} = 1\) and \(\Bat{a}{\itellp} = \B{0}\).
In this case, the function \(f(\Bat{c}{\itellp})\) would
have the same value as in the previous iteration:
\[
    \begin{aligned}
    f(\Bat{c}{\itellp}) &= \left\|H+\Bat{c}{\itell}
    (\Bat{c}{\itell})^\T\right\|^2_\F \\
    &= \|H\|^2_\F - \left(\lambda_{\min}((\at{Q}{\itell})^\T H
        \at{Q}{\itell})\right)^2 \\
    & < \|H\|^2_\F.
    \end{aligned}
\]
This demonstrates that the objective function value must decrease
in each iteration. Additionally, in each iteration, we must have
\(\lambda_{\min}((\at{Q}{\itellp})^\T H \at{Q}{\itellp}) < 0\).

\section{Ground-State Energy Computation Time of \ch{N2} Using Different Basis Sets and Different Number of Threads}
\label{app:scalability}

Tables~\ref{tab:scalability-vdz} to~\ref{tab:scalability-vqz}
present the mCDFCI computation times for the \ch{N2} molecule across
varying number of threads, using the cc-pVDZ, cc-pVTZ, and cc-pVQZ
basis sets, respectively. These results are already illustrated
in Figure~\ref{fig:scalability}.
Here, we provide the
corresponding parallel efficiency and a detailed time breakdown
for each computational step.

\reviewerone{The time required for variable updates decreases steadily
as the number of threads increases, demonstrating the strong scalability
of the multi-threaded implementation.
Conversely, the other two non-parallelized components initially
exhibit a decrease in time, followed by an eventual increase.
The initial decrease can be attributed to the reduced number of iterations
required as the number of coordinates grows, given that when \(k\) is small,
the time to solve a \((k+1) \times (k+1)\)  eigenvalue problem
remains relatively constant.
However, as the number of coordinates continues to increase,
the \(k^3\) scaling in the eigenvalue computation becomes increasingly
significant, resulting in a noticeable rise in the time required
for local energy updates. To address this bottleneck,
parallelizing these components represents an important area
for future development of the CDFCI software package.}

\begin{table*}[h!]
    \centering
    \begin{tabular}{c|cc|>{\revieweronecolor}c >{\revieweronecolor}c >{\revieweronecolor}c}
    \toprule
    \textbf{Threads} & \textbf{Time (s)} & \textbf{Parallel} & \textbf{Variable} & \textbf{Local Energy} & \textbf{Others} \\
    & & \textbf{Efficiency} & \textbf{Update (s)} & \textbf{Update (s)} & \textbf{(s)} \\
    \midrule
    \midrule
    \(1\) & \(2076.40\) & \(1.000\) & \(1974.46\) & \(12.28\) & \(89.29\) \\
    \(2\) & \(1038.48\) & \(1.000\) & \(980.75\) & \(7.67\) & \(49.66\) \\
    \(4\) & \(505.32\) & \(1.027\) & \(474.03\) & \(4.78\) & \(26.33\) \\
    \(8\) & \(365.66\) & \(0.710\) & \(345.40\) & \(3.84\) & \(16.26\) \\
    \(16\) & \(192.09\) & \(0.676\) & \(178.04\) & \(3.85\) & \(10.11\) \\
    \(32\) & \(106.10\) & \(0.612\) & \(92.77\) & \(4.94\) & \(8.34\) \\
    \(64\) & \(65.91\) & \(0.492\) & \(49.03\) & \(6.99\) & \(9.82\) \\
    \(128\) & \(89.31\) & \(0.182\) & \(28.07\) & \(44.65\) & \(16.54\) \\
    \(256\) & \(195.71\) & \(0.041\) & \(24.09\) & \(138.11\) & \(33.45\) \\
    \bottomrule
    \end{tabular}
    \caption{Time, Parallel Efficiency, \reviewerone{and Component Breakdown} for Ground-State Energy of \ch{N2} Using cc-pVDZ (14e, 28o)}
    \label{tab:scalability-vdz}
\end{table*}

\begin{table*}[h!]
    \centering
    \begin{tabular}{ccc>{\revieweronecolor}c>{\revieweronecolor}c>{\revieweronecolor}c}
    \toprule
    \textbf{Threads} & \textbf{Time (s)} & \textbf{Parallel} & \textbf{Variable} & \textbf{Local Energy} & \textbf{Others} \\
    & \textbf{} & \textbf{Efficiency} & \textbf{Update (s)} & \textbf{Update (s)} & \textbf{(s)} \\
    \midrule
    \(1\) & \(16626.01\) & \(1.000\) & \(16074.56\) & \(22.67\) & \(528.26\) \\
    \(2\) & \(7003.45\) & \(1.187\) & \(6720.30\) & \(11.92\) & \(270.99\) \\
    \(4\) & \(3084.88\) & \(1.347\) & \(2939.85\) & \(7.61\) & \(137.25\) \\
    \(8\) & \(2045.62\) & \(1.016\) & \(1967.34\) & \(5.10\) & \(73.07\) \\
    \(16\) & \(1045.78\) & \(0.994\) & \(1001.69\) & \(4.36\) & \(39.65\) \\
    \(32\) & \(531.91\) & \(0.977\) & \(502.56\) & \(5.42\) & \(23.86\) \\
    \(64\) & \(272.70\) & \(0.953\) & \(247.74\) & \(7.39\) & \(17.50\) \\
    \(128\) & \(192.42\) & \(0.675\) & \(125.78\) & \(45.50\) & \(21.09\) \\
    \(256\) & \(289.45\) & \(0.224\) & \(91.93\) & \(152.79\) & \(44.68\) \\
    \bottomrule
    \end{tabular}
    \caption{Time, Parallel Efficiency, \reviewerone{and Component Breakdown} for Ground-State Energy of \ch{N2} Using cc-pVTZ (14e, 60o).}
    \label{tab:scalability-vtz}
\end{table*}

\begin{table*}[h!]
    \centering
    \begin{tabular}{ccc>{\revieweronecolor}c>{\revieweronecolor}c>{\revieweronecolor}c}
    \toprule
    \textbf{Threads} & \textbf{Time (s)} & \textbf{Parallel} & \textbf{Variable} & \textbf{Local Energy} & \textbf{Others} \\
    & \textbf{} & \textbf{Efficiency} & \textbf{Update (s)} & \textbf{Update (s)} & \textbf{(s)} \\
    \midrule
    \(1\) & \(53182.43\) & \(1.000\) & \(51319.96\) & \(36.06\) & \(1825.70\) \\
    \(2\) & \(22897.89\) & \(1.161\) & \(21927.33\) & \(26.10\) & \(943.89\) \\
    \(4\) & \(9834.53\) & \(1.352\) & \(9347.02\) & \(10.28\) & \(477.05\) \\
    \(8\) & \(6669.42\) & \(0.997\) & \(6358.72\) & \(9.77\) & \(300.77\) \\
    \(16\) & \(3190.37\) & \(1.042\) & \(3031.49\) & \(8.08\) & \(150.68\) \\
    \(32\) & \(1485.38\) & \(1.119\) & \(1393.64\) & \(6.90\) & \(84.75\) \\
    \(64\) & \(699.20\) & \(1.188\) & \(610.01\) & \(8.84\) & \(80.29\) \\
    \(128\) & \(524.26\) & \(0.793\) & \(371.17\) & \(43.86\) & \(109.17\) \\
    \(256\) & \(475.64\) & \(0.437\) & \(245.80\) & \(142.78\) & \(87.00\) \\
    \bottomrule
    \end{tabular}
    \caption{Time, Parallel Efficiency, \reviewerone{and Component Breakdown} for Ground-State Energy of \ch{N2} Using cc-pVQZ (14e, 110o).}
    \label{tab:scalability-vqz}
\end{table*}

\section{\ch{N2} Binding Curve Data for Bond Length and Energy}
    \label{app:dissociation}
    Table~\ref{tab:dissociation} shows the binding curve data
    for four sets of calculations: Hartree--Fock (HF),
    CDFCI with the cc-pVDZ basis, CDFCI with the full cc-pVQZ (110 orbitals),
    and CDFCI with a reduced set of 28 compressed orbitals derived from
    the cc-pVQZ basis using the OptOrbFCI method~\cite{Li2020}.
    The energies for HF were generated using the PSI4 package~\cite{psi4}
    version 1.8. The energies for the cc-pVDZ and cc-pVQZ (28)
    basis sets are cited from previous studies~\cite{Wang2019, Li2020}.
    The table includes the bond lengths in atomic units (\(a_0\))
    and the corresponding ground-state energies in Hartree (Ha).

    \begin{table*}[h!]
        \begin{tabular}{cccccc}
        \toprule
        \textbf{Bond Length ($a_0$)} & \textbf{Hartree--Fock} & \textbf{cc-pVDZ} & \textbf{cc-pVQZ (28)} & \textbf{cc-pVQZ (110)} \\
        \midrule
        \( 1.50 \)  & \( -108.484166 \) & \( -108.630048 \) & \( -108.814403 \) & \( -108.910666 \) \\
        \( 1.55 \)  & \( -108.607446 \) & \( -108.771997 \) & \( -108.939182 \) & \( -109.036535 \) \\
        \( 1.60 \)  & \( -108.707310 \) & \( -108.888846 \) & \( -109.042905 \) & \( -109.139272 \) \\
        \( 1.65 \)  & \( -108.787292 \) & \( -108.984314 \) & \( -109.126015 \) & \( -109.222403 \) \\
        \( 1.70 \)  & \( -108.850397 \) & \( -109.061575 \) & \( -109.192655 \) & \( -109.288927 \) \\
        \( 1.75 \)  & \( -108.899180 \) & \( -109.123348 \) & \( -109.244370 \) & \( -109.341398 \) \\
        \( 1.80 \)  & \( -108.935822 \) & \( -109.171964 \) & \( -109.283011 \) & \( -109.381998 \) \\
        \( 1.85 \)  & \( -108.962179 \) & \( -109.209426 \) & \( -109.313701 \) & \( -109.412578 \) \\
        \( 1.90 \)  & \( -108.979837 \) & \( -109.237458 \) & \( -109.335975 \) & \( -109.434729 \) \\
        \( 1.95 \)  & \( -108.990152 \) & \( -109.257541 \) & \( -109.351156 \) & \( -109.449809 \) \\
        \( 2.00 \)  & \( -108.994283 \) & \( -109.270953 \) & \( -109.360360 \) & \( -109.458975 \) \\
        \( 2.05 \)  & \( -108.993221 \) & \( -109.278790 \) & \( -109.364582 \) & \( -109.463219 \) \\
        \( 2.10 \)  & \( -108.987816 \) & \( -109.281994 \) & \( -109.364756 \) & \( -109.463348 \) \\
        \( 2.15 \)  & \( -108.978797 \) & \( -109.281374 \) & \( -109.361496 \) & \( -109.460105 \) \\
        \( 2.20 \)  & \( -108.966786 \) & \( -109.277621 \) & \( -109.355553 \) & \( -109.454133 \) \\
        \( 2.25 \)  & \( -108.952320 \) & \( -109.271328 \) & \( -109.347340 \) & \( -109.445965 \) \\
        \( 2.30 \)  & \( -108.935857 \) & \( -109.263001 \) & \( -109.337391 \) & \( -109.436048 \) \\
        \( 2.35 \)  & \( -108.917789 \) & \( -109.253072 \) & \( -109.325979 \) & \( -109.424753 \) \\
        \( 2.40 \)  & \( -108.898453 \) & \( -109.241908 \) & \( -109.313533 \) & \( -109.412410 \) \\
        \( 2.45 \)  & \( -108.878135 \) & \( -109.229823 \) & \( -109.297016 \) & \( -109.399302 \) \\
        \( 2.50 \)  & \( -108.857081 \) & \( -109.217083 \) & \( -109.283533 \) & \( -109.385670 \) \\
        \( 2.60 \)  & \( -108.813564 \) & \( -109.190508 \) & \( -109.255799 \) & \( -109.357597 \) \\
        \( 2.70 \)  & \( -108.769211 \) & \( -109.163600 \) & \( -109.227940 \) & \( -109.329427 \) \\
        \( 2.80 \)  & \( -108.724949 \) & \( -109.137358 \) & \( -109.200871 \) & \( -109.302021 \) \\
        \( 2.90 \)  & \( -108.681424 \) & \( -109.112473 \) & \( -109.175147 \) & \( -109.275997 \) \\
        \( 3.00 \)  & \( -108.639072 \) & \( -109.089405 \) & \( -109.151280 \) & \( -109.251749 \) \\
        \( 3.10 \)  & \( -108.598176 \) & \( -109.068450 \) & \( -109.129515 \) & \( -109.229566 \) \\
        \( 3.20 \)  & \( -108.558912 \) & \( -109.049779 \) & \( -109.102007 \) & \( -109.209619 \) \\
        \( 3.30 \)  & \( -108.521376 \) & \( -109.033462 \) & \( -109.084817 \) & \( -109.191894 \) \\
        \( 3.40 \)  & \( -108.485608 \) & \( -109.019484 \) & \( -109.070067 \) & \( -109.176446 \) \\
        \( 3.50 \)  & \( -108.451609 \) & \( -109.007747 \) & \( -109.057557 \) & \( -109.163193 \) \\
        \( 3.60 \)  & \( -108.419352 \) & \( -108.998083 \) & \( -109.047199 \) & \( -109.152036 \) \\
        \( 3.70 \)  & \( -108.388793 \) & \( -108.990269 \) & \( -109.038765 \) & \( -109.142780 \) \\
        \( 3.80 \)  & \( -108.359873 \) & \( -108.984050 \) & \( -109.031901 \) & \( -109.135188 \) \\
        \( 3.90 \)  & \( -108.332527 \) & \( -108.979162 \) & \( -109.026548 \) & \( -109.129007 \) \\
        \( 4.00 \)  & \( -108.306685 \) & \( -108.975357 \) & \( -109.022348 \) & \( -109.123980 \) \\
        \( 4.10 \)  & \( -108.282276 \) & \( -108.972410 \) & \( -109.018986 \) & \( -109.119960 \) \\
        \( 4.20 \)  & \( -108.259228 \) & \( -108.970132 \) & \( -109.016448 \) & \( -109.116684 \) \\
        \( 4.30 \)  & \( -108.237469 \) & \( -108.968366 \) & \( -109.014495 \) & \( -109.114082 \) \\
        \( 4.40 \)  & \( -108.216928 \) & \( -108.966991 \) & \( -109.012920 \) & \( -109.111935 \) \\
        \( 4.50 \)  & \( -108.197538 \) & \( -108.965910 \) & \( -109.011722 \) & \( -109.110256 \) \\
        \bottomrule
        \end{tabular}
        \caption{Binding curve data for bond length and energy,
        showing the results from Hartree--Fock (HF), cc-pVDZ,
        and CDFCI with both the full cc-pVQZ basis (110 orbitals)
        and the compressed cc-pVQZ basis (28 orbitals).}
        \label{tab:dissociation}
    \end{table*}

\clearpage
\begin{figure*}[th]
    \centering
    \textbf{\Large TOC Graphic}\\[1em] 
    \includegraphics[width=0.8\textwidth]{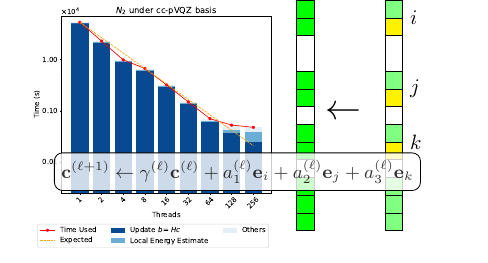}
\end{figure*}

\end{document}